\documentclass{werner_smac}
\usepackage{bm,cite,amsmath,amssymb,graphicx,float}
\usepackage{listings} \newcommand{\SMAC}{\textsf{SMAC}}

\newcommand{\forward}{\text{fw}} % forward
\newcommand{\backward}{\text{bw}} % backward
\newcommand{\mc}{1,1} % coupling time
\newcommand{\taucoup}{\tau_{\text{coup}}} % coupling time
\newcommand{\taucorr}{\tau_{\text{corr}}} % coupling time
\usepackage[small,bf]{wwwww_caption}
\usepackage{enumerate}
\title{Four lectures\\ on\\ computational statistical physics}
\author{Werner Krauth\thanks{Lectures given at the 2008 Les Houches Summer School \quot{Exact methods in low-dimensional physics and quantum computing}. }}
\affiliation{Laboratoire de Physique Statistique\\CNRS-Ecole Normale
Sup\'{e}rieure \\24 rue Lhomond\\
75231 Paris Cedex 05}
\begin{document}
%
% definition of float: algorithm
%
\newfloat{algorithm}{ht}{loa}
\floatname{algorithm}{{\bf Algorithm}}
\floatname{figure}{{\bf Fig.}}
\newcommand{\graybar}{\noindent\fcolorbox{black}{gray}{\parbox{309pt}{\mbox{}}}}
%
% Definitions for the constants pi and i (i^2 =-1)
%
\newcommand{\mpi}{\pi}
\newcommand{\mi}{\mathrm{i}}
%
% Definitions for text, equation numbers, figures, etc
%
\newcommand{\clabel}[1]{\label{c:#1}}
\newcommand{\elabel}[1]{\label{e:#1}}
\newcommand{\exlabel}[1]{\label{ex:#1}} % Exercice label
\newcommand{\slabel}[1]{\label{s:#1}}
\newcommand{\alabel}[1]{\label{a:#1}}
\newcommand{\eq}[1]{eqn~(\ref{e:#1})}
\newcommand{\eqq}[1]{Equation~(\ref{e:#1})}
\newcommand{\EQ}[1]{(\ref{e:#1})}
\newcommand{\eqtwo}[2]{eqns~(\ref{e:#1}) and~(\ref{e:#2})}
\newcommand{\EQTWO}[2]{Equations~(\ref{e:#1}) and~(\ref{e:#2})}
\newcommand{\fig}[1]{Fig.~\ref{f:#1}}
\newcommand{\figtwo}[2]{Figs~\ref{f:#1} and~\ref{f:#2}}
\newcommand{\figg}[1]{Figure~\ref{f:#1}}
\newcommand{\FIG}[1]{\ref{f:#1}}
\newcommand{\quot}[1]{``#1''}
\newcommand{\tab}[1]{Table~\ref{t:#1}} 
\newcommand{\TAB}[1]{\ref{t:#1}} 
\newcommand{\chap}[1]{Chapter~\ref{c:#1}} 
\newcommand{\sect}[1]{Section~\ref{s:#1}} 
\newcommand{\SECT}[1]{\ref{s:#1}} 
\newcommand{\subsect}[1]{Subsection~\ref{s:#1}} 
\newcommand{\SUBSECT}[1]{\ref{s:#1}} 
\newcommand{\ex}[1]{Exerc.~\ref{ex:#1}} 
\newcommand{\EX}[1]{\ref{ex:#1}} 
\newcommand{\eg}{\textrm{e.g.}}
\newcommand{\cf}{\textrm{cf}}
\newcommand{\etc}{\textrm{etc.}}
\newcommand{\etcp}{\textrm{etc}}
\newcommand{\ie}{\textrm{i.e.}}
\newcommand{\vs}{\textrm{vs.}}
\newcommand{\prog}[1]{Alg.~\sub{#1}}
\newcommand{\PROG}[1]{\ref{a:#1} (\sub{#1})}
\newcommand{\progg}[1]{Algorithm~\ref{a:#1} (\sub{#1})}
\newcommand{\proggg}[1]{Algorithm \sub{#1}}
\newcommand{\progn}[2]{Alg.~\ref{a:#1_#2}}
\newcommand{\prognn}[2]{\ref{a:#1_#2}}
\newcommand{\mylap}{year 2005 laptop computer}
\newcommand{\sub}[1]{\texttt{#1}}
\newcommand{\perm}[2]{\left(\begin{smallmatrix}#1\\#2\end{smallmatrix}\right)}
\newcommand{\fin}{\bf $^*$}
\newcommand{\oo}{\cdot}
%
% text-indices
%
\newcommand{\accept}{{\text{accept}}}  %  acceptance rate, etc
\newcommand{\anti}{{\text{a}}}  %  antisymmetric  (wave function, etc)
\newcommand{\betac}{{\beta_{\text{c}}}}  %  Critical inverse temperature
\newcommand{\btm}{\text{btm}} %  bosonic trap model
\newcommand{\cube}{{\text{cube}}}  %  cut-off
\newcommand{\cut}{{\text{cut}}}  %  cut-off
\newcommand{\Ekin}{E_{\text{kin}}}  %  Boltzmann constant
\newcommand{\free}{{\text{free}}}  %  free (hamiltonian, etc)
\newcommand{\harm}{{\text{h.o.}}}  %  harmonic (hamiltonian, etc)
\newcommand{\kB}{k_{\text{B}}}  %  Boltzmann constant
\newcommand{\lab}{\text{lab}}  %  laboratory system
\newcommand{\Ll}{\text{l}}  %  large particle (capitalized because \large is already taken)
\newcommand{\Ss}{\text{s}}  %  small particle (capitalized because \small is already taken)
\newcommand{\new}{{\text{new}}}  %  new
\newcommand{\old}{{\text{old}}}  %  old
\newcommand{\pair}{{\text{pair}}}  %  rejection rate, etc
\newcommand{\reject}{{\text{reject}}}  %  rejection rate, etc
\newcommand{\rel}{{\text{rel}}}  %  for 'relative'
\newcommand{\resc}{\text{resc}}  % rescaled
\newcommand{\rot}{\text{rot}}  %  rotating system
\newcommand{\sat}{\text{sat}}  %  sat (saturation density, etc)
\newcommand{\symm}{{\text{s}}}  %  symmetric (wave function, etc)
\newcommand{\Tc}{{T_{\text{c}}}}  %  Critical temperature
\newcommand{\Tr}{\text{Tr}}  %  Trace
\newcommand{\test}{{\text{test}}}  %  test value...
\newcommand{\tot}{{\text{tot}}}  %  total number
\newcommand{\Var}[1]{\text{Var}\lc #1 \rc}
\newcommand{\wall}{{\text{wall}}}  %  wall (collision time)
%
%  non-diagonal density matrix , with optional parameter for `sym' `box' `cube' etc
%
\newcommand{\rhomat}[4][]{\rho^{\text{#1}\!}\lc #2,#3,#4\rc}
\newcommand{\todo}{[\dots]}
\newcommand{\TODO}[1]{
\begin{center}
\framebox{
\begin{minipage}{\columnwidth}
#1
\end{minipage}}
\end{center}}
%
% new definitions for figures and mfigures, including labels and captions, 
% allows easy interchange 
%
% parameter #1: name of figure (without .eps), serves as label; #2: caption

\newcommand{\smacfigure}[2]{\begin{figure}[htbp]
\begin{center}
\includegraphics{Figures/#1.eps}
\end{center}
\caption
{#2}
\label{f:#1}
\end{figure}}

\newcommand{\smaccfigure}[3][0.5]{
\vspace{#1 cm}
   \begin{cfigure}[htbp]{f:#2}
\cfigparts{\parentdir/Illustrations/#2.eps}
{#3}
\end{cfigure}}

\newcommand{\smactfigure}[3]{\begin{figure}[htbp]
\begin{center}
\includegraphics{Figures/#1.eps}
\includegraphics{Figures/#2.eps}
\end{center}
\caption
{#3}
\label{f:#1}
\end{figure}}

\newcommand{\smacmfigure}[3][0pt]{\mfigure[#1]{\begin{center}\includegraphics
{\parentdir/Illustrations/#2.eps}\end{center}
\mfigcaption{#3}
\label{f:#2}
}}  

\newcommand{\smacpfigure}[2]{
\exfigure{\parentdir/Illustrations/#1.eps}
\exfigcaption*{f:#1}{#2}}

\newcommand{\smactable}[4]{
\begin{table}[htbp]
\tableparts
{
\caption{#4}
\label{t:#1}
}
{\begin{tabular}{#2}
\hline
#3
\hline
\end{tabular}}
\end{table}}

\newcommand{\smacmtable}[5][0pt]{
\mtable[#1]{\mtableparts{
\mtabcaption
{#5}
\label{t:#2}
}
{\begin{tabular}{#3}
\hline
#4
\hline
\end{tabular} }} }

\newcommand{\problemsection}[1]{\subsubsection{(\sect{#1})}}
%
% representation of floating point numbers
%
\newcommand{\fpn}[2]{\ensuremath{#1 \! \times \! 10^{#2}}}
%
% definition of constants which always have to be the same: Nsites Ntrial ndim
% dos
%
%
%  abbreviations for matcal
%
\newcommand{\NCAL}{\mathcal{N}}  %  mathcal
\newcommand{\OCAL}{\mathcal{O}}  %  mathcal
%
% the condition command 
%
\newcommand{\cond}[1]{\Theta \left \{ #1 \right \}  }
%
% the common Greek letters and their bold versions
%
\newcommand{\s}{\sigma}
\newcommand{\stilde}{\tilde{\sigma}}
\newcommand{\bs}{\boldsymbol{\sigma}}
\newcommand{\eps}{\epsilon}
%
% poor man's bold for + and - (used for spins)
%
%\newcommand{\minus}{\ensuremath{\boldsymbol{-}}}
%\newcommand{\plus}{\ensuremath{\boldsymbol{+}}}
\newcommand{\minus}{\ensuremath{\pmb{-}}}
\newcommand{\plus}{\ensuremath{\pmb{+}}}
\newcommand{\eqntext}[1]{ \left\{\!\!\!\mbox{\begin{tabular}{c}  #1 \end{tabular}}\!\!\! \right\}}
%
% Definitions for algorithms
%
\newcommand{\PROCEDURE}[1]{\textbf{procedure}\ \sub{#1}}
\newcommand{\CALL}[2][]{\textbf{call}\ \FUNCTION[#1]{#2} }
\newcommand{\FUNCTION}[2][]{\sub{#2} \glb #1 \grb }
\newcommand{\BRACE}[1]{
\;\;\;  \left\{\begin{array}{l}#1\end{array} \right.}
\newcommand{\IS}[2]{#1 \leftarrow #2}
\newcommand{\SWAP}[2]{#1 \leftrightarrow #2}
\newcommand{\FOR}[1]{\textbf{for}\ #1\ \textbf{do}}
\newcommand{\WHILE}[1]{\textbf{while (} #1 \textbf{) do}}
\newcommand{\ENDPROCEDURE}{\text{------} \\ \vspace{-0.8cm}}
\newcommand{\ENDPROCEDURET}{\text{------} \\ \vspace{-0.5cm}} % ENDPROCEDURE IN TEXT
\newcommand{\GOTO}[1]{\textbf{goto}\ #1}
\newcommand{\CONTINUE}{\textbf{continue}}
\newcommand{\DO}{\textbf{do}}
\newcommand{\EXIT}{\textbf{exit}}
\newcommand{\IF}[1]{\textbf{if (} #1 \textbf{)}}
\newcommand{\IFTHEN}[1]{\textbf{if (} #1 \textbf{) then}}
\newcommand{\ELSEIFTHEN}[1]{\textbf{else if (} #1 \textbf{) then}}
\newcommand{\IFELSE}[3]{\textbf{if (} #1 \textbf{) \{} #2 \textbf{\} else \{} #3 \textbf{\}} }
\newcommand{\AND}{\text{and}}
\newcommand{\OR}{\text{or}}
\newcommand{\SETPLUS}{\uplus}  % used in      SET \SETPLUS {j}
\newcommand{\THEN}{\textbf{then}}
\newcommand{\STOP}{\textbf{stop}}
\newcommand{\ELSE}{\textbf{else}}
\newcommand{\ELSEIF}[1]{\textbf{else if (} #1 \textbf{) then}}
\newcommand{\OUTPUT}[1]{\textbf{output}\ #1}
\newcommand{\SORT}[1]{\text{sort[} #1\text{]}}
\newcommand{\INPUT}[1]{\textbf{input}\ #1}
\newcommand{\COMMENT}[1]{ \text{\footnotesize (#1)}}
\newcommand{\TABLE}[1]{\{#1\}}
\newcommand{\LIST}[1]{\{#1\}} 
\newcommand{\NULLLIST}{\LIST{0 \TO 0 }} 
\newcommand{\SET}[1]{\{#1\}}
\newcommand{\EMPTYSET}{\{ \}}
\newcommand{\NULLSET}{\SET{0 \TO 0 }} 
%
% Definitions for display math
%
\newcommand{\BRA}[1]{\ensuremath{\langle #1 |}} % Dirac notation 
\newcommand{\BRAN}[1]{\ensuremath{\langle #1 }} % Dirac notation  Bra without |
\newcommand{\KET}[1]{\ensuremath{|#1 \rangle }} % Dirac notation
\newcommand{\exph}[1]{\mathrm{e}^{#1}} % high exponential (not in programs)
\newcommand{\exphp}[1]{\sub{e}^{#1}}  % high exponential (in programs) (should not be used)
\newcommand{\expl}[1]{\exp \lb #1 \rb } % low exponential (not in programs)
\newcommand{\explp}[1]{\sub{exp} \lb  #1 \rb} % exponential (in programs)
%
% exponentials with braces 'a': no brace 'b' () 'c' [] 'd' {}
%
\newcommand{\expa}[1]{\mathrm{e}^{#1}}   % high exponential groupings a
\newcommand{\expb}[1]{\exp \glb #1 \grb} % low exponential with groupings b
\newcommand{\expba}[1]{\exp \bigl( #1 \bigr)} % low exponential with groupings b-small
\newcommand{\expbb}[1]{\exp \Bigl( #1 \Bigr)} % low exponential with groupings b-medium
\newcommand{\expbc}[1]{\exp \biggl( #1 \biggr)} % low exponential with groupings b-large
\newcommand{\expbd}[1]{\exp \Biggl( #1 \Biggr)} % low exponential with groupings b-Xlarge
\newcommand{\expc}[1]{\exp \glc #1 \grc} % low exponential with groupings c
\newcommand{\expd}[1]{\exp \gld #1 \grd} % low exponential with groupings d
%
% trigonometric functions with braces 'a': no brace 'b' () 'c' [] 'd' {}
%
\newcommand{\sina}[2][]{\sin^{#1} \! \gla #2 \gra}  % sin-brace,  with - nothing
\newcommand{\cosa}[2][]{\cos^{#1} \! \gla #2 \gra}  % cos-brace,  with - nothing
\newcommand{\tana}[2][]{\tan^{#1} \!\gla #2 \gra}  % tan-brace,  with - nothing
\newcommand{\cota}[2][]{\cot^{#1} \!\gla #2 \gra}  % cot-brace,  with - nothing
\newcommand{\sinha}[2][]{\sinh^{#1}\! \gla #2 \gra} % sinh-brace, with - nothing
\newcommand{\cosha}[2][]{\cosh^{#1}\! \gla #2 \gra} % cosh-brace, with - nothing
\newcommand{\tanha}[2][]{\tanh^{#1}\! \gla #2 \gra} % tanh-brace, with - nothing
\newcommand{\cotha}[2][]{\coth^{#1}\! \gla #2 \gra} % coth-brace, with - nothing

\newcommand{\sinb}[2][]{\sin^{#1} \glb #2 \grb}  % sin-brace,  with - ()
\newcommand{\cosb}[2][]{\cos^{#1} \glb #2 \grb}  % cos-brace,  with - ()
\newcommand{\tanb}[2][]{\tan^{#1} \glb #2 \grb}  % tan-brace,  with - ()
\newcommand{\cotb}[2][]{\cot^{#1} \glb #2 \grb}  % cot-brace,  with - ()
\newcommand{\sinhb}[2][]{\sinh^{#1} \glb #2 \grb} % sinh-brace, with - ()
\newcommand{\coshb}[2][]{\cosh^{#1} \glb #2 \grb} % cosh-brace, with - ()
\newcommand{\tanhb}[2][]{\tanh^{#1} \glb #2 \grb} % tanh-brace, with - ()
\newcommand{\cothb}[2][]{\coth^{#1} \glb #2 \grb} % coth-brace, with - ()

\newcommand{\sinc}[2][]{\sin^{#1} \glc #2 \grc}  % sin-brace,  with - []
\newcommand{\cosc}[2][]{\cos^{#1} \glc #2 \grc}  % cos-brace,  with - []
\newcommand{\tanc}[2][]{\tan^{#1} \glc #2 \grc}  % tan-brace,  with - []
\newcommand{\cotc}[2][]{\cot^{#1} \glc #2 \grc}  % cot-brace,  with - []
\newcommand{\sinhc}[2][]{\sinh^{#1} \glc #2 \grc} % sinh-brace, with - []
\newcommand{\coshc}[2][]{\cosh^{#1} \glc #2 \grc} % cosh-brace, with - []
\newcommand{\tanhc}[2][]{\tanh^{#1} \glc #2 \grc} % tanh-brace, with - []
\newcommand{\cothc}[2][]{\coth^{#1} \glc #2 \grc} % coth-brace, with - []

\newcommand{\sind}[2][]{\sin^{#1} \gld #2 \grd}  % sin-brace,  with - {}
\newcommand{\cosd}[2][]{\cos^{#1} \gld #2 \grd}  % cos-brace,  with - {}
\newcommand{\tand}[2][]{\tan^{#1} \gld #2 \grd}  % tan-brace,  with - {}
\newcommand{\cotd}[2][]{\cot^{#1} \gld #2 \grd}  % cot-brace,  with - {}
\newcommand{\sinhd}[2][]{\sinh^{#1} \gld #2 \grd} % sinh-brace, with - {}
\newcommand{\coshd}[2][]{\coth^{#1} \gld #2 \grd} % cosh-brace, with - {}
\newcommand{\tanhd}[2][]{\tanh^{#1} \gld #2 \grd} % tanh-brace, with - {}
\newcommand{\cothd}[2][]{\coth^{#1} \gld #2 \grd} % coth-brace, with - {}

\newcommand{\loga}[2][]{\log^{#1}\! \gla #2 \gra}  % log-brace,  with - nothing
\newcommand{\logb}[2][]{\log^{#1} \glb #2 \grb}  % log-brace,  with - ()
\newcommand{\logc}[2][]{\log^{#1} \glc #2 \grc}  % log-brace,  with - []
\newcommand{\logd}[2][]{\log^{#1} \gld #2 \grd}  % log-brace,  with - {}

\newcommand{\arccosa}[2][]{\arccos^{#1}\! \gla #2 \gra}  % arccos-brace,  with - nothing
\newcommand{\arccosb}[2][]{\arccos^{#1} \glb #2 \grb}  % arccos-brace,  with - ()
\newcommand{\arccosc}[2][]{\arccos^{#1} \glc #2 \grc}  % arccos-brace,  with - []
\newcommand{\arccosd}[2][]{\arccos^{#1} \gld #2 \grd}  % arccos-brace,  with - {}

\newcommand{\arctana}[2][]{\arctan^{#1}\! \gla #2 \gra}  % arctan-brace,  with - nothing
\newcommand{\arctanb}[2][]{\arctan^{#1} \glb #2 \grb}  % arctan-brace,  with - ()
\newcommand{\arctanc}[2][]{\arctan^{#1} \glc #2 \grc}  % arctan-brace,  with - []
\newcommand{\arctand}[2][]{\arctan^{#1} \gld #2 \grd}  % arctan-brace,  with - {}

\newcommand{\gl}{\left(}  % ' group left' 
\newcommand{\gr}{\right)}  % ' group right' 
\newcommand{\lb}{\left[}  % left brace `[' used in sinh[3 beta] etc
\newcommand{\rb}{\right]}  % right brace ']' "           " 
\newcommand{\lc}{\left(}  % left brace `(' used in groupings of expressions
\newcommand{\rc}{\right)}  % right brace ')' "         "            " 
\newcommand{\mult}{\times} % multiplication sign in display math, 
\newcommand{\multcc}{\cdot} % multiplication: character times character
\newcommand{\multnn}{\times} % multiplication:    number times number 
\newcommand{\multcn}{\cdot} % multiplication: character times number
\newcommand{\ran}[2]{\FUNCTION[#1,#2]{ran}}
\newcommand{\nran}[2]{\FUNCTION[#1,#2]{nran}}
\newcommand{\perL}{\text{per,$L$}}
\newcommand{\perLL}{\text{per,$2L$}}
\newcommand{\boxx}[1]{\text{box,$#1$}}

%
% Groupings on levels a--nothing, b--(), c--[], and d--{} (better solution)
%
\newcommand{\gla}{\,}  % ' group left a' 
\newcommand{\gra}{}  % ' group right a' 
\newcommand{\glb}{\left(}  % ' group left b' 
\newcommand{\grb}{\right)}  % ' group right b' 
\newcommand{\glc}{\left[}  % ' group left c'
\newcommand{\grc}{\right]}  % ' group right c' 
\newcommand{\gld}{\left\{}  % ' group left d' 
\newcommand{\grd}{\right\}}  % ' group right d' 

\newcommand{\const}{\text{const}}
\newcommand{\Nhits}{N_{\text{hits}}}
\newcommand{\gauss}[1]{\FUNCTION[#1]{gauss}}
\newcommand{\PLUSPLUS}{+ \dots +}
\newcommand{\PLUSDOTS}{+ \cdots}
\newcommand{\MINUSPLUS}{- \dots +}
\newcommand{\MINUSDOTS}{- \cdots}
\newcommand{\PLUSMINUS}{+ \dots -}
\newcommand{\MINUSMINUS}{- \dots -}
\newcommand{\MULTMULT}{\cdot \cdots \cdot}
\newcommand{\TIMESTIMES}{\times \cdots \times}
\newcommand{\TO}{,\ldots,}
\newcommand{\VEC}[1]{\mathbf{#1}}
%
% vector notations
%
\newcommand{\avec}{\VEC{a}}
\newcommand{\Avec}{\VEC{A}}
\newcommand{\bvec}{\VEC{b}}
\newcommand{\cvec}{\VEC{c}}
\newcommand{\Cvec}{\VEC{C}}
\newcommand{\dvec}{\VEC{d}}
\newcommand{\evec}{\VEC{e}}
\newcommand{\ehatvec}{\hat{\VEC{e}}}
\newcommand{\fvec}{\VEC{f}}
\newcommand{\Fvec}{\VEC{F}}
\newcommand{\gvec}{\VEC{g}}
\newcommand{\hvec}{\VEC{h}}
\newcommand{\ivec}{\VEC{i}}
\newcommand{\jvec}{\VEC{j}}
\newcommand{\kvec}{\VEC{k}}
\newcommand{\lvec}{\VEC{l}}
\newcommand{\mvec}{\VEC{m}}
\newcommand{\nvec}{\VEC{n}}
\newcommand{\ovec}{\VEC{o}}
\newcommand{\pvec}{\VEC{p}}
\newcommand{\qvec}{\VEC{q}}
\newcommand{\rvec}{\VEC{r}}
\newcommand{\Rvec}{\VEC{R}}
\newcommand{\svec}{\VEC{s}}
\newcommand{\tvec}{\VEC{t}}
\newcommand{\uvec}{\VEC{u}}
\newcommand{\vvec}{\VEC{v}}
\newcommand{\wvec}{\VEC{w}}
\newcommand{\xvec}{\VEC{x}}
\newcommand{\Xvec}{\VEC{X}}
\newcommand{\yvec}{\VEC{y}}
\newcommand{\Yvec}{\VEC{Y}}
\newcommand{\zvec}{\VEC{z}}
\newcommand{\Zvec}{\VEC{Z}}
\newcommand{\atilde}{\tilde{a}}
\newcommand{\btilde}{\tilde{b}}
\newcommand{\ctilde}{\tilde{c}}
\newcommand{\ptilde}{\tilde{p}}
\newcommand{\vtilde}{\tilde{v}}
\newcommand{\xtilde}{\tilde{x}}
\newcommand{\ytilde}{\tilde{y}}
\newcommand{\ztilde}{\tilde{z}}
\newcommand{\Etilde}{\tilde{E}}
\newcommand{\Rtilde}{\tilde{R}}
\newcommand{\phitilde}{\tilde{\phi}}
\newcommand{\Psitilde}{\tilde{\Psi}}
\newcommand{\psitilde}{\tilde{\psi}}
\newcommand{\sigmatilde}{\tilde{\sigma}}
\newcommand{\alphavec}{\boldsymbol{\alpha}}
\newcommand{\betavec}{\boldsymbol{\beta}}
\newcommand{\gammavec}{\boldsymbol{\gamma}}
\newcommand{\deltavec}{\boldsymbol{\delta}}
\newcommand{\pivec}{\boldsymbol{\pi}}
\newcommand{\epsvec}{\boldsymbol{\epsilon}}
\newcommand{\sigmavec}{\boldsymbol{\sigma}}
\newcommand{\del}{\delta}
\newcommand{\dd}[1]{\text{d}{#1\ }}   % this is for differentials in formulas
\newcommand{\ddd}[1]{\text{d}{#1}}   % same as above, without space 
%
% Delta_tau Delta_x Delta_y, etc
%
\newcommand{\Delbeta}{\Delta_{\beta}}
\newcommand{\DelE}{\Delta_E}
\newcommand{\Deli}{\Delta_i}
\newcommand{\Delk}{\Delta_k}
\newcommand{\Dellambda}{\Delta_{\lambda}}
\newcommand{\Delmu}{\Delta_{\mu}}
\newcommand{\DelM}{\Delta_{M}}
\newcommand{\Deln}{\Delta_n}
\newcommand{\Delphi}{\Delta_{\phi}}
\newcommand{\Delr}{\Delta_r}
\newcommand{\Delt}{\Delta_t}
\newcommand{\Deltau}{\Delta_{\tau}}
\newcommand{\Delvvec}{\Delta_{\vvec}}
\newcommand{\Delx}{\Delta_x}
\newcommand{\Delxp}{\Delta'_x}
\newcommand{\Delxk}{\Delta_{x_k}}
\newcommand{\Delxvec}{\Delta_{\xvec}}
\newcommand{\Delxvecp}{\Delta_{\xvec'}}
\newcommand{\Dely}{\Delta_y}
\newcommand{\Delyvec}{\Delta_{\yvec}}
\newcommand{\Delz}{\Delta_z}

%
% scalar product
%
\newcommand{\neigh}[2]{\langle #1 , #2 \rangle}
\newcommand{\scal}[2]{(#1 \pmb{\cdot} #2)}
\newcommand{\mean}[1]{\left\langle #1 \right\rangle}
\newcommand{\half}{\frac{1}{2}}
\newcommand{\thalf}{\tfrac{1}{2}}
\newcommand{\edge}[2]{(#1,#2)}
%
% References
%
\newcommand{\tit}[5]{(#4) #5,\ {\it #1}\ {\bf #2}, #3}
\newcommand{\titnv}[4]{(#3) #4,\ {\it #1}, #2}
\newcommand{\titprep}[4]{(#3) #4,\ {#1}, #2}
\newcommand{\titbook}[4]{(#3) {\it #4}, \ {#1}, {#2}}
%
%
% Journal references
%
\def\epl{Europhysics Letters}
\def\jap{Journal of Applied Physics}
\def\jpco{J.\ Phys.\ Cond.\ Mat.\ }
\def\jpc{J.\ Phys.\ C}
\def\jpa{Journal of Physics A}
\def\jpsj{J.\ Phys.\ Soc.\ Jpn.}
\def\jsp{Journal of Statistical Physics}
\def\pr{Physical Review}
\def\pra{Physical Review A}
\def\prb{Physical Review B}
\def\pre{Physical Review E}
\def\prl{Physical Review Letters}
\def\rmp{Reviews of Modern Physics}
\def\zpb{Z.\ Phys.\ B}
\def\sci{Science}
\def\natu{Nature}
\def\jcp{Journal of Chemical Physics}
\def\cjp{Canadian  Journal of Physics}
\def\ijmpb{International Journal of Modern Physics B}
\def\jmp{Journal of Mathematical Physics}
% end definitions
%
% Index-related routines
%
%\newcommand{\algindex}{\rule{.5mm}{2mm}\index}
%\newcommand{\authindex}{\rule{.5mm}{2mm}\index}
%\newcommand{\subindex}{\rule{.5mm}{2mm}\index}
\newcommand{\algindex}{\index}
\newcommand{\authindex}{\index}
\newcommand{\subindex}{\index}
\newcommand{\progi}[2]{\sub{#2} (Alg. \ref{a:#1_#2})}
\maketitle
%s1 ###
\preface
In my lectures at the Les Houches Summer School 2008, I discussed central
concepts of computational statistical physics, which I felt would be
accessible to the very cross-cultural audience at the school.

I started with a discussion of sampling, which lies at the heart
of the Monte Carlo approach. I specially emphasized the concept of
perfect sampling, which offers a synthesis of the traditional direct and
Markov-chain sampling  approaches. The second lecture concerned classical
hard-sphere systems, which illuminate  the foundations of statistical
mechanics, but also illustrate the curious difficulties that beset even
the most recent simulations. I then moved on, in the third lecture,
to quantum Monte Carlo methods, that underly much of the modern work in
bosonic systems. Quantum Monte Carlo is an intricate subject. Yet one can
discuss it in simplified settings (the single-particle free propagator,
ideal bosons) and write direct-sampling algorithms for the two cases
in two or three dozen lines of code only. These idealized algorithms
illustrate many of the crucial ideas in the field.  The fourth lecture
attempted to illustrate aspects of the unity of physics as realized in
the Ising model simulations of recent years.

More details on what I discussed in Les Houches, and wrote up (and
somewhat rearranged) here, can be found in my book, \quot{Statistical
Mechanics: Algorithms and Computations} (\SMAC), as well
as in recent papers. Computer programs are available for download and
perusal at the book's web site www.smac.lps.ens.fr.

\tableofcontents
\maintext

\chapter{Sampling}
\section{Direct sampling, sample transformation}
\slabel{direct_sample_sample_transformation}

As an illustration of what is meant by sampling, and how it relates
to integration, we consider the Gaussian integral:
\begin{equation}
\int_{-\infty}^{\infty} \frac{\ddd{x}}{\sqrt{2 \mpi}} \expb{-x^2/2} = 1.
\elabel{IN_error_integral}
\end{equation}

This integral can be computed by taking its square:
\begin{align}
\left[\int_{-\infty}^{\infty} \frac{\ddd{x}}{\sqrt{2 \mpi}} \expb{-x^2/2} \right]^2 & =
\int_{-\infty}^{\infty} \frac{\ddd{x}}{\sqrt{2 \mpi}} \exph{-x^2/2}
\int_{-\infty}^{\infty} \frac{\ddd{y}}{\sqrt{2 \mpi}} \exph{-y^2/2} \elabel{first_of_gauss} \\
\ldots & = \int_{-\infty}^{\infty} \frac{\ddd{x} \ddd{y}}{2 \mpi} \expc{-(x^2+y^2)/2},
\intertext{and then switching to polar coordinates ($\ddd{x} \ddd{y} = r \ddd{r} \ddd{\phi}$), }
\ldots & = \int_{0}^{2 \mpi} \frac{\ddd{\phi}}{2 \mpi} \int_{0}^{\infty} r \dd{r} \expb{-r^2/2}, 
\intertext{as well as performing the substitutions $r^2/2=\Upsilon$ ($r \ddd{r} = \ddd{\Upsilon}$) and $\expb{-\Upsilon} = \Psi$}
\ldots & = \int_{0}^{2 \mpi} \frac{\ddd{\phi}}{2 \mpi}\
\int_{0}^{\infty} \dd{\Upsilon} \expb{-\Upsilon}\\
\ldots & = \int_{0}^{2 \mpi} \frac{\ddd{\phi}}{2 \mpi}\
\int_{0}^{1} \dd{\Psi} = 1.
\elabel{IN_gauss_eqd}
\end{align}
In our context, it is less important  that we can do the integrals in
\eq{IN_gauss_eqd} analytically, than that $\phi$ and $\Psi$ can be sampled as uniform random
variables in the interval $[0,2 \pi]$ (for $\phi$) and in $[0,1]$
(for $\Psi$). Samples $\phi=\ran{0}{2\mpi}$ and $\Psi = \ran{0}{1}$, are
readily obtained from the random number generator $\ran{a}{b}$ lingering
on any computer. We can plug these random numbers into the substitution
formulas which took us from \eq{first_of_gauss} to \eq{IN_gauss_eqd},
and that take us now from two uniform random numbers $\phi$ and $\Psi$
to Gaussian random numbers $x$ and $y$.  We may thus apply the
integral transformations in the above equation to the samples, in other
words perform a \quot{sample transformation}.  This is a practical
procedure for generating Gaussian random numbers from uniform random
numbers, and we best discuss it as what it is, namely an algorithm.
\begin{algorithm}
\newcommand{\algo}{gauss}
\begin{center}
$ \begin{array}{l}
\PROCEDURE{\algo}\\
\IS{\phi}{\ran{0}{2 \mpi}}\\
\IS{\Psi}{\ran{0}{1}}\\
\IS{\Upsilon}{-\sub{log}\ \Psi}\\
\IS{r}{\sqrt{2 \Upsilon}}\\
\IS{x}{r \sub{cos}\ \phi}\\
\IS{y}{r \sub{sin}\ \phi}\\
\OUTPUT{\SET{x,y}}\\
\ENDPROCEDURE\
\end{array} $
\caption{\sub{\algo}. \SMAC\ pseudocode program for transforming
two uniform random numbers $\phi, \Psi$ into two independent uniform
Gaussian random numbers $x,y$. Except for typography, \SMAC\ pseudocode
is close to the Python programming language.}
\alabel{\algo}
\end{center}
\end{algorithm}

\SMAC\ pseudocode can be implemented in many computer languages
\footnote{The \SMAC\ web page www.smac.lps.ens.fr provides programs in
language ranging from Python, Fortran, C, and Mathematica to TI basic,
the language of some pocket calculators.}. Of particular interest is the
computer language Python, which resembles pseudocode, but is executable
on a computer exactly as written. We continue in these lectures using
and showing Python code, as in algorithm 1.1.

\begin{lstlisting}[caption=gausstest.py]
from random import uniform as ran, gauss
from math import sin, cos, sqrt, log, pi
def gausstest(sigma):
  phi = ran(0,2*pi)
  Upsilon = -log(ran(0.,1.)) 
  r = sigma*sqrt(2*Upsilon)
  x = r*cos(phi)
  y = r*sin(phi)
  return x,y
print gausstest(1.)
\end{lstlisting}

Direct-sampling algorithms exist for arbitrary one-dimensional
distributions (see \SMAC\ Sect. 1.2.4). Furthermore, arbitrary discrete
distributions $\SET{\pi_1 \TO \pi_K}$ can be directly sampled, after an
initial effort of about  $K$ operations with $\sim \log_2 K$ operation
by \quot{Tower sampling} (see \SMAC\ Sect. 1.2.3). This means that
sampling a distribution made up of, say, one billion terms takes only
about 30 steps. We will use this fact in \sect{landsberg_sampling},
for the direct sampling algorithm for ideal bosons.  Many   trivial
multi-dimensional distribution (as for example non-interacting particles)
can also be sampled.

Direct-sampling algorithms
also solve much less trivial problems as for example the free path
integral, ideal bosons, and the two-dimensional Ising model, in fact, many
problems which possess an exact analytic solution.  These direct-sampling
algorithms are often the race-car engines inside general-purpose
Markov-chain algorithms for complicated interacting problems.

Let us discuss the computation of integrals derived from the sampled
ones $Z = \int \dd{x} \pi(x)$ (in our example, the distribution $\pi(x)$
is the Gaussian)
\begin{equation*}
\frac{
\int \dd{x} \OCAL(x) \pi(x) }
{
\int \dd{x}\pi(x)} \simeq \frac{1}{N} \sum_{i=1}^{N} \OCAL(x_i), 
\end{equation*}
where the points $x_i$ are sampled from the distribution $\pi(x)$.
This approach, as shown, allows to compute mean values of observables for
a given distribution $\pi$. One can also bias the
distribution function $\pi$ using Markov-chain approaches. This gives
access to a large class of very non-trivial distributions.

\section{Markov-chain sampling}
\slabel{Markov_chain_sampling}
Before taking up the discussion of elaborate systems (hard spheres,
bosons, spin glasses), we first concentrate on a single particle in a
finite one-dimensional lattice $k = 1 \TO N$ (see \fig{one_d_single}).
This case is even simpler than the aforementioned Gaussian, because the
space is discrete rather than continuous, and because the
site-occupation probabilities $\SET{\pi_1 \TO \pi_N}$ are all the same
\begin{equation*}
\pi_k = \frac{1}{N}\quad \forall k.
\elabel{equal_probability}
\end{equation*}
We can sample this trivial distribution by picking $k$ as a random integer
between $1$ and $N$. Let us nevertheless study a Markov-chain algorithm,
whose diffusive dynamics converges towards the probability distribution
of \eq{equal_probability}.  For concreteness, we consider the algorithm
where at all integer times $t \in ]-\infty, \infty[$, the particle hops
with probability $\tfrac{1}{3}$ from one site to each of its neighbors:
\begin{equation}
  p_{k \to k+1} = p_{k \to k-1} = 1/3\quad\text{(if possible)}.
  \elabel{algo_probabilities}
\end{equation}
The probabilities to remain on the site are $1/3$ in the interior and,
at the boundaries,  $p_{1 \to 1} = p_{N \to N}=2/3$.

\smacfigure{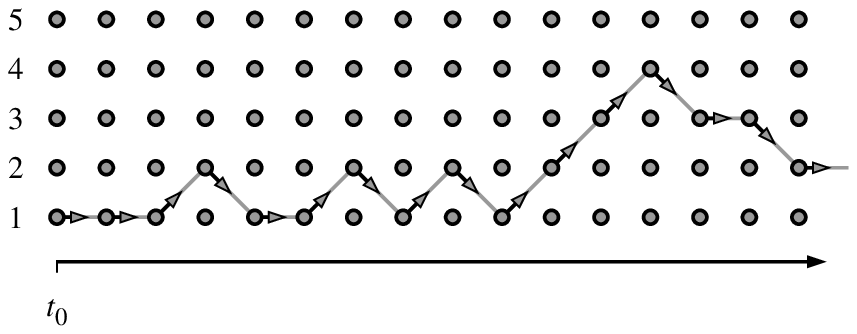}
{Markov-chain algorithm on a five-site lattice. A single particle 
hops towards a site's neighbor
with probability $1/3$, and remains on the site with probability $1/3$
(probability $2/3$ on the boundary).}

These transition probabilities satisfy the notorious detailed balance condition
\begin{equation}
\pi_k p_{k\to l} = \pi_l p_{l \to k}
\elabel{detailed_balance}
\end{equation} 
for the constant probability distribution $\pi_k$. Together with the ergodicity
of the algorithm this  guarantees that in the infinite-time limit, the
probability $\pi_k$ to find the particle at site $k$ is indeed independent
of $k$.  With appropriate transition probabilities,  the Metropolis
algorithm would allow us to sample any generic distribution $\pi_k$ or $\pi(x)$
in one or more dimensions (see \SMAC\ Sect. 1.1.5).

Markov-chain methods are more general than direct approaches, but the
price to pay is that they converge to  the target distribution $\pi$
only in the infinite-time limit. The nature of this  convergence can
be analyzed by the transfer matrix $T^{\mc}$ (the matrix of transition
probabilities, in our case a $5\times 5$ matrix):
\begin{equation}
T^{\mc} =
\{p_{ k \to l} \} = \frac{1}{3}
\begin{pmatrix}
2 & 1 & 0 & 0 & 0 \\
1 & 1 & 1 & 0 & 0 \\
0 & 1 & 1 & 1 & 0 \\
0 & 0 & 1 & 1 & 1 \\
0 & 0 & 0 & 1 & 2 \\
\end{pmatrix}
\elabel{}
\end{equation}
The eigenvalues of the transfer matrix $T^{\mc}$ are $(1,\half \pm
\sqrt{5}/6, \frac{1}{6} \pm \sqrt{5}/6)$. The largest eigenvalue, equal
to one, expresses the conservation of the sum of all probabilities
$\sum(\pi_1 \PLUSPLUS \pi_5)=1$.  Its corresponding eigenvector is
$|\bullet \circ \circ \circ \circ \rangle + \cdots + |\circ \circ \circ
\circ \bullet \rangle $, by construction, because of the detailed-balance
condition \eq{detailed_balance}. The second-largest eigenvalue, $\lambda^{\mc}_2 = \half +
\sqrt{5}/6 = 0.8727$ governs the decay of correlation functions at
large times. This is easily seen by computing the probability vector
$\pivec(t_0 + \tau) = \SET{\pi_1(t) \TO \pi_5(t)}$ which can be written
in terms of the eigenvalues and eigenvectors

Let us suppose that the simulations always start\footnote{we should
assume that the initial configuration is different from the stationary
solution because otherwise all the coefficients $\alpha_2 \TO \alpha_N$
would be zero.} on site $k=1$, and let us decompose this initial configuration 
onto the eigenvectors $\pivec(0) = \alpha_1\pivec^{e}_1 \PLUSPLUS  \alpha_N \pivec^{e}_N$.
\begin{equation*}
\pivec(t) = \gl T^{\mc}\gr^t \pivec(0) = \pivec^{e}_1 + \alpha_2 \lambda_2^t \pivec^e_2 + \dots
\elabel{}
\end{equation*}
At large times, corrections to the equilibrium state vanish
as $\lambda_2^{\tau}$, so that site probabilities approach the equilibrium 
value as $\expb{\tau/\taucorr}$ with a time constant
\begin{equation}
\taucorr =1/ |\loga{\lambda_2}| \simeq 7.874
\elabel{correlation_time}
\end{equation}
(see \SMAC\ Sect. 1.1.4, p. 19f). 

We retain that the exponential convergence of a Monte Carlo algorithm
is characterized by a scale, the convergence time $\taucorr$. We also
retain that convergence takes place after a few $\taucorr$, in our example
say $3 \times 7.87 \sim 25$ iterations (there is absolutely no need to
simulate for an infinite time). If our  Markov chains always start at
$t=0$  on the site $1$, it is clear that for all times $t' < \infty$, the
occupation probability $\pi_1(t')$ of  site $1$ is larger than, say, the
probability $\pi_5(t')$ to be on site $5$. This would make us believe that
Markov chain simulations never completely decorrelate from the initial
condition, and are always somehow less good than direct-sampling. This
belief is wrong, as we shall discuss in the next section.

\section{Perfect sampling}
\slabel{perfect_sampling}
We need  to simulate for no more than a few $\taucorr$, but we must
not stop our calculation short of this time.  This is the critical
problem in many real-life situations, where we cannot compute the
correlation time as reliably as in our five-site problem: $\taucorr$
may be much larger than we suspect, because the empirical approaches for
determining  correlation times may have failed (see \SMAC\ Sect. 1.3.5).
In contrast, the problem of the exponentially small tail 
for $\tau \gg \taucorr$  is totally irrelevant.

Let us take up again our five-site problem,  with the goal of obtaining
rigorous information about convergence from within the simulation.
As illustrated in \fig{one_d_diffusion}, the Markov-chain algorithm can
be formulated in terms of time sequences of random maps: Instead
of prescribing one move per time step, as we did in \fig{one_d_single},
we now sample moves independently for all sites $k$ and each time $t$.
At time $t_0$, for example, the particle should move straight from sites
$1$, $2$ and $5$ and down from sites $3$ and $4$, etc.  Evidently, for
a single particle, there is no difference between the two formulations,
and detailed balance is satisfied either way.  With the formulation
in terms of random maps, we can verify that from time $t_0+\taucoup$
on, all initial conditions generate the same output.  This so-called
\quot{coupling} is of great interest because after the coupling time
$\taucoup$, the influence of the initial condition has completely
disappeared.  In the rest of this lecture, we consider extended Monte
Carlo simulations as the one in \fig{one_d_diffusion}, with arrows drawn
for each site and time.

\smacfigure{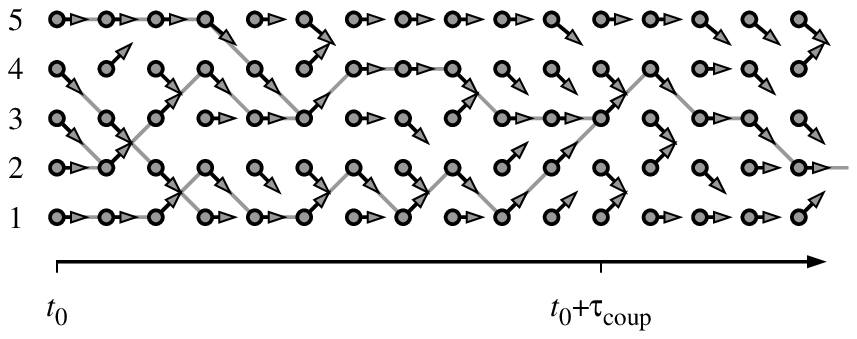} {Extended Monte Carlo simulation on $N=5$
sites. In this example, the coupling time is $\taucoup = 11$.}

The coupling time $\taucoup$ is a random variable ($\taucoup=11$
in \fig{one_d_diffusion}) whose distribution $\pi(\taucoup)$
vanishes exponentially in the limit $\taucoup \to \infty$ because the
random maps at different times are independent.

The extended Monte Carlo dynamics describes a physical system
with from $1$ to $N$ particles, and transition probabilities, from
\eq{algo_probabilities}, as for example:
\begin{equation}
\begin{aligned}
p^\forward(|\circ \circ \bullet \circ \bullet   \rangle & \to |\circ \bullet \circ \bullet \circ\rangle ) = 1/9\\
p^\forward(|\circ \circ \bullet \circ \bullet   \rangle & \to |\circ \circ \bullet \circ \bullet\rangle ) = 2/9\\
p^\forward(|\circ \circ \bullet \circ \bullet   \rangle & \to |\circ \circ \circ \bullet   \circ\rangle ) = 1/9\\
p^\forward(|\circ \circ \bullet \bullet \bullet \rangle & \to |\circ \circ \circ \bullet \circ\rangle ) = 1/27, 
\end{aligned}
\elabel{forward_probabilities}
\end{equation}
etc., We may analyze the convergence of this system through its transfer
matrix $T^{\forward}$ (in our case, a $31\times 31$ matrix, because of the $31$
non-empty states on five sites):
\begin{equation*}
T^\forward =
\begin{pmatrix}
T^{1,1} & T^{2,1} & \dots & \dots \\
   0 & T^{2,2} & T^{3,2}& \dots \\
   0 & 0   & T^{3,3}& \dots \\
\dots \\
   0 & 0   &    & T^{N,N} \\
\end{pmatrix}
\elabel{}
\end{equation*}
where the block $T^{k,l}$ connects states with $k$ particles at time
$t$ to states with $l\le k$ particles at time $t+1$.  The matrix
$T^\forward$ has an eigenvector with eigenvalue $1$ which describes
the equilibrium, and a second-largest eigenvalue which describes the
approach to equilibrium, and yields the coupling time $\taucoup$.
Because of the block-triangular structure of $T^\forward$, the
eigenvalues of $T^{1,1}$ constitute the single-particle sector of
$T^\forward$, including the largest eigenvalue $\lambda^\forward_1=1$,
with corresponding left eigenvector $|\bullet\circ\circ\circ\circ\rangle
+ \cdots + |\circ\circ\circ\circ \bullet\rangle$. The second-largest
eigenvalue of $T^\forward$ belongs to the $(2,2)$ sector. It is given by
$\lambda^{\forward}_2 = 0.89720 > \lambda_2^{\text{MC}}$, and it describes
the behavior of the coupling probability $\pi(\taucoup)$ for large times.

Markov chains shake off all memory of their initial conditions at
time $\taucoup$, but this time changes from simulation to simulation:
it is a random variable.  Ensemble averages over (extended) simulations
starting at $t_0$, and ending at $t_0 + \tau$ thus contain mixed averages
over coupled and \quot{not yet coupled} runs, and only the latter carry
correlations with the initial condition. To reach pure ensemble averages
over coupled simulations only, one has to start the simulations at time $t
= -\infty$, and go up to $t=0$.  This procedure, termed \quot{coupling
from the past} (Propp and Wilson 1996), is familiar to theoretical
physicists because in dynamical calculations, the initial condition is
often put to $-\infty$. Let us now see how this trick can be put to use
in practical calculations.

\smacfigure{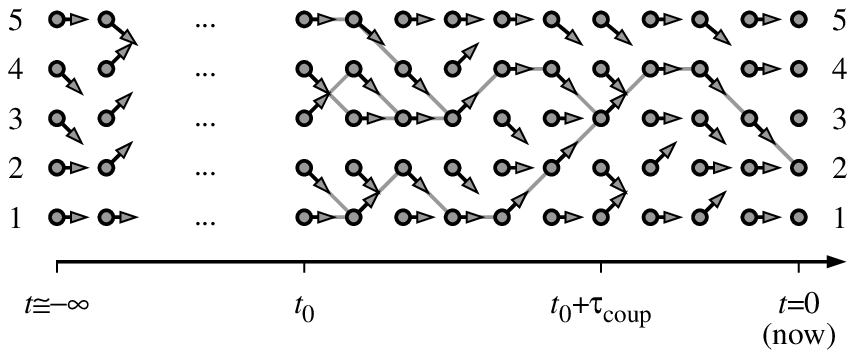} {Extended simulation on $N=5$ sites. The outcome
of this infinite simulation, from $t=-\infty$ through $t=0$, is $k=2$.}

An extended simulation \quot{from the past} is shown in \fig{one_d_cftp}.
It has run for an infinite time so that all the eigenvalues of the transfer
matrix but $\lambda_1=1$ have died away and only the equilibrium state
has survived.  The configuration at $t=0$ is thus a perfect sample, and
each value $k$ ($ k \in \SET{1\TO 5}$ is equally likely, if we average
over all extended Monte Carlo simulations.  For the specific simulation
(choice of arrows in \fig{one_d_cftp}), we know that the simulation must pass
by one of the five points at time $t=t_0=-10$).  However,
the Markov chain couples between $t=t_0$ and $t=0$, so that we know that
$k(t=0)=2$. If it did not couple at $t=t_0$, we would have to provide
more information (draw more arrows, for $t=t_0-1, t_0 -2, \dots$)
until the chain couples.

Up to now, we have considered the forward transfer matrix, but there is
also a backward matrix $T^\backward$, which describes the propagation of
configurations from $t=0$ back to $t=-1$, $t=-2$, etc.  $A^\backward$ is
similar to the matrix $A^\forward\sim A^\backward$ ( the two matrices
have identical eigenvalues), because the probability distribution to
couple between time $t=-|t_0|$ and $t=0$ equals the probability
to couple between time $t=0$ and $t = +|t_0|$. This implies that the
distributions of coupling times in the forward and backward directions
are identical.

We have also seen that the eigenvalue corresponding to the coupling time
is larger than the one for the correlation time.  For our one-dimensional
diffusion problem, this can be proven exactly. More generally, this is
because connected correlation functions decay as
\begin{equation*}
\mean{\OCAL(0) \OCAL(t)}_c \sim \expa{-t /\taucorr}. 
\elabel{}
\end{equation*}
Only the non-coupled (extended) simulations contribute
to this correlation function, and their proportion is equal to
$\expb{-t/\taucoup}$. We arrive at
\begin{equation*}
\expc{-t\glb  1/\taucoup -1/\taucorr\grb}  \sim \mean{\OCAL(0) \OCAL(t)}_c^{\text{non-coup}}, 
\elabel{}
\end{equation*}
and $\taucoup > \taucorr$ because even the non-coupling correlation
functions should decay with time.

Finally, we have computed in \fig{one_d_diffusion} and in \fig{one_d_cftp}
the coupling time by following all possible initial conditions. This
approach is unpractical for more complicated problems,
such as the Ising model on $N$ sites with its $2^N$ configurations. Let us
show in the five-site model how this problem can sometimes be overcome:
It suffices to define a new Monte Carlo dynamics in which trajectories
cannot cross, by updating, say, even lattice sites ($k=2,4$) at every
other time $t_0, t_0+2, \dots$ and odd lattice sites ($k=1,3,5$) at
times $t_0+1, t_0+3, \dots$. The coupling of the trajectories starting
at sites $1$ and $5$ obviously determines $\taucoup$.

\smacfigure{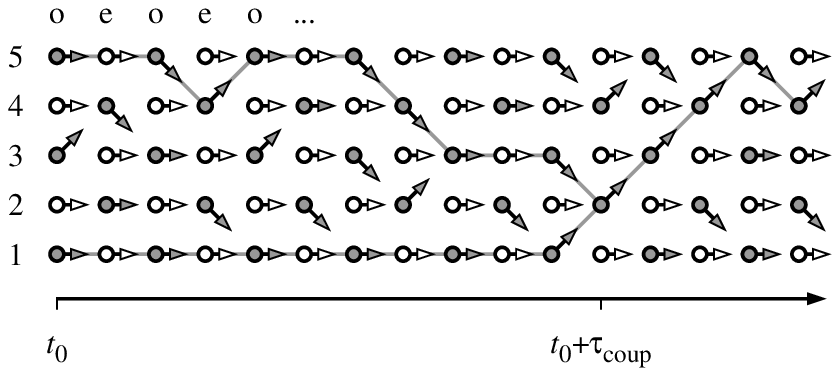} {Extended Monte Carlo simulation with
odd and even times. Trajectories cannot cross, and the coupling of the
two extremal configurations (with $k(t_0)=1$ and $k(t_0=5)$) determines
the coupling time.}

In the Ising model, the ordering relation of \fig{one_d_alternating}
survives in the form of a \quot{half-order} between configurations (see
\SMAC\ Sect. 5.2.2), but in the truly interesting models, such as spin
glasses, no such trick is likely to exist. One must really supervise
the $2^N$ configurations at $t=t_0$.  This non-trivial task has been
studied extensively (see Chanal and Krauth 2008).

\chapter{Classical hard-sphere systems}

\section{Molecular dynamics}
\slabel{molecular_dynamics}
Before statistical mechanics, not so long ago, there only was classical
mechanics. The junction between the two has fascinated generations of
mathematicians and physicists, and nowhere can it be studied better than
in the hard-sphere system: $N$ particles in a box, moving about like
billiard balls, with no other interaction than the hard-sphere
exclusion (without friction or angular momentum).  For more than
a century, the hard-sphere model has been a prime example for
statistical mechanics and a parade ground for rigorous mathematical
physics. For more than fifty years, it has served as a test bed for
computational physics, and it continues to play this role.

For concreteness, we first consider four disks (two-dimensional spheres)
in a square box with walls (no periodic boundary conditions), as in
\fig{event_movie_double}: From an initial configuration, as the one
at $t=0$, particles fly apart on straight trajectories either until one
of them hits a wall or until two disks undergo an elastic collision. The
time for this next \quot{event} can be computed exactly by going over
all disks (taken by themselves, in the box, to compute the time for the
next wall collisions) and all pairs of disks (isolated from the rest of
the system, and from the walls, to determine the next pair collisions),
and then treating the event closest in time (see \SMAC\ Sect. 2.1).

\smacfigure{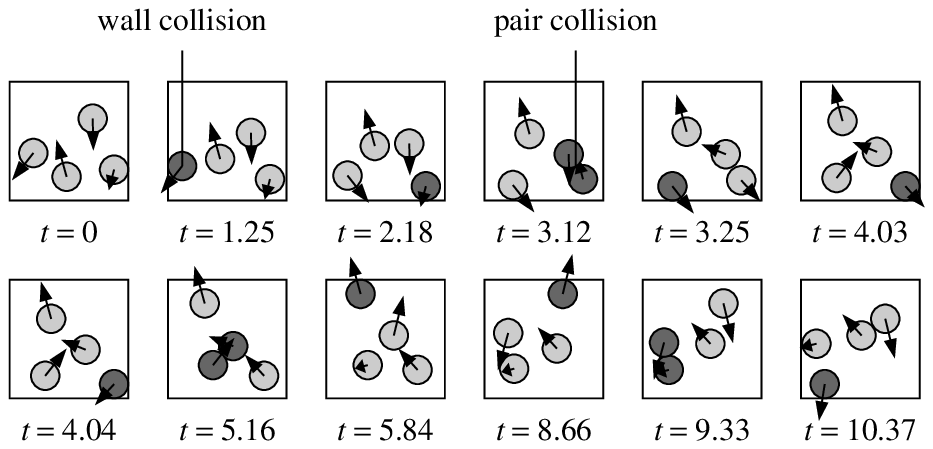}
{Event-driven molecular dynamics simulation with four hard disks in a
square box.}
The event-chain algorithm can be implemented in a few dozen lines,
just a few too many for a free afternoon in Les Houches
(program listings are available on the \SMAC\ web site). It implements the entire
dynamics of the $N$-particle system without time discretization, because
there is no differential equation to be solved. The only error committed stems
from the finite-precision arithmetic implemented on a computer.

This error manifests itself surprisingly quickly, even in our simulation
of four disks in a square: typically, calculations done in $64$bit
precision ($15$ significant digits) get out of step with other calculations from identical
initial conditions with $32$bit calculations after a few dozen pair
collisions.  This is the manifestation of chaos in the hard-sphere system.
The appearance of numerical uncertainties for given initial conditions
can be delayed by using even higher precision arithmetic, but it is out of
the question to control a calculation that has run for a few minutes on
our laptop, and gone through a few billion collisions.
Chaos in the hard sphere model has its origin in the negative
curvature of the sphere surfaces, which magnifies tiny differences
in the trajectory at each pair collision and causes serious roundoff
errors in computations\footnote{as it causes humiliating experiences
at the billiard table}.  

The mathematically rigorous analysis of chaos in the hard-disks system
has met with resounding success: Sinai (1970) (for two disks) and Simanyi
(2003, 2004) (for general hard-disk and hard-sphere systems) were able
to mathematically prove the foundations of statistical physics for the
system at hand, namely the equiprobability principe
for hard spheres: This means that during an infinitely long  molecular
dynamics simulation,  the probability density satisfies the following:
\begin{equation*}
\eqntext{probability of configuration with\\
$[\xvec_1,\xvec_1+ \ddd{\xvec_1}] \TO
[\xvec_N,\xvec_N + \ddd{\xvec_N}]$}
\propto \pi(\xvec_1 \TO \xvec_N)\, \ddd{\xvec_1} \TO \ddd{\xvec_N},
\end{equation*}
where
\begin{equation}
\pi(\xvec_1 \TO \xvec_N) =
\begin{cases}
1  & \text{if configuration legal ($|\xvec_k - \xvec_l| > 2 \sigma$ for $k \ne l$)}\\
0  & \text{otherwise}\\
\end{cases}.
\elabel{direct-diskdens}
\end{equation}

An analogous property has been proven for the velocities: 
\begin{equation}
\pi(\vvec_1 \TO \vvec_N) =
\begin{cases}
1  & \text{if velocities legal ($\sum \vvec_k^2 = E_{\text{kin}}/(2m))$}\\
0  & \text{otherwise}\\
\end{cases}.
\elabel{direct-disk_velocity}
\end{equation}
Velocities are legal if they add up to the correct value of the conserved
kinetic energy, and their distribution is constant on the surface of
the $2N$ dimensional hypersphere of radius $\sqrt{E_{\text{kin}}/(2m)}$
(for disks).

The two above equations contain all of equilibrium statistical physics
in a nutshell. The equal-probability principle of \eq{direct-diskdens}
relates to the principle that two configurations of the same energy have
the same statistical weight. The sampling problem for velocities, in
\eq{direct-disk_velocity} can be solved with Gaussians, as discussed
in \SMAC\ Sect. 1.2.6.  It reduces to the Maxwell distribution for large
$N$ (see \SMAC\ Sect. 2.2.4) and implies the Boltzmann distribution
(see \SMAC\ Sect. 2.3.2).

\section{Direct sampling, Markov chain sampling}
\slabel{direct_sampling_hard_spheres}

We now move on from molecular dynamics simulations to the Monte Carlo sampling.
To sample $N$ disks with the constant probability distribution of \eq{direct-diskdens},
we uniformly throw a set of $N$ particle positions $\SET{\xvec_1 \TO \xvec_N}$
into the square. Each of these sets of $N$ positions is generated with
equal probability. We then sort out all those sets that are no legal
hard-sphere configurations. The remaining ones (the gray configurations
in \fig{direct_movie}) still have equal probabilities, exactly as
called for in \eq{}.
\smacfigure{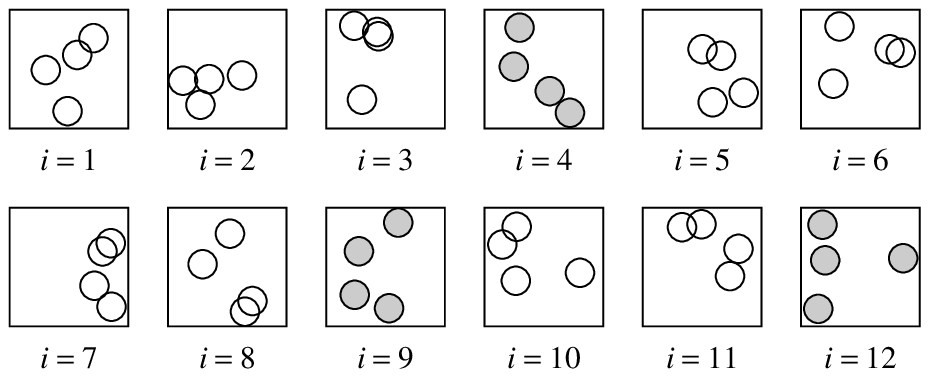}
{Direct-sampling Monte Carlo algorithm for hard disks in a box without
periodic boundary conditions (see \prog{direct-disks.py}).}
In the Python programming language, we can implement this algorithm in
a few lines (see \prog{direct-disks.py}): one places up to $N$ particles
at random positions (see line 7 of \prog{direct-disks.py}). If two disks
overlap, one breaks out of this construction and restarts with an empty
configuration.  The rejection rate $p_{\text{accept}}(N,\sigma)$ of this
algorithm (the probability to generate legal (gray) configurations in
\fig{direct_movie}):
\begin{equation*}
p_{\text{accept}} = \frac{\text{number of valid configurations with radius $\sigma$}}
{\text{number of valid configurations with radius $0$}}= Z_\sigma/Z_0
\elabel{}
\end{equation*}
is exponentially small both in the particle
number $N$ and in the density of particles, and this for physical reasons
(see the discussion in \SMAC\ Sect. 2.2.2)).
\begin{lstlisting}[float,caption=direct-disks.py]
from random import uniform as ran
sigma=0.20
condition = False
while condition == False:
  L = [(ran(sigma,1-sigma),ran(sigma,1-sigma))]
  for k in range(1,4): # 4 particles considered
   b = (ran(sigma,1-sigma),ran(sigma,1-sigma))
   min_dist = min((b[0]-x[0])**2+(b[1]-x[1])**2 for x in L)
   if min_dist < 4*sigma**2:
     condition = False
     break
   else:
     L.append(b)
     condition = True
print L
\end{lstlisting}

For the hard-sphere system, Markov-chain methods are much more widely
applicable than the direct-sampling algorithm. In order to satisfy the
detailed balance condition of \eq{detailed_balance}, we must impose
that the probability to move from a legal configuration $a$ to another,
$b$, must be the same as the probability to move from $b$ back to $a$
(see \fig{markov_disks_movie}). This is realized most easily by
picking a random disk and moving it inside a small square around its
original position, as implemented in \prog{markov-disks.py}.

\smacfigure{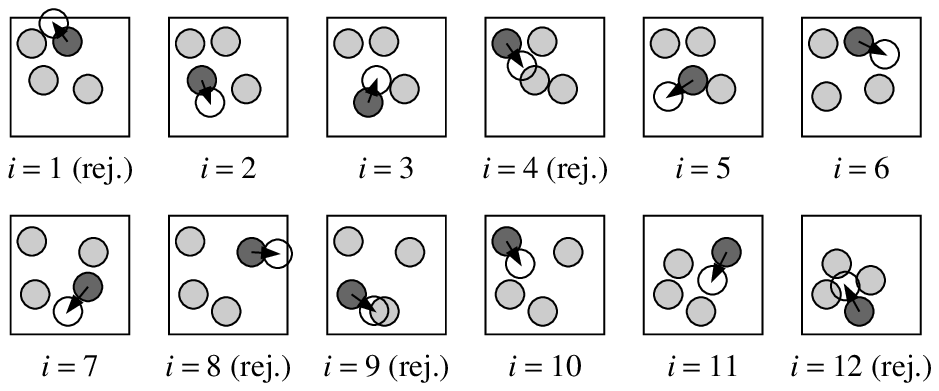}
{Markov-chain Monte Carlo algorithm for hard disks in a box without
periodic boundary conditions.}

\begin{lstlisting}[float,caption=markov-disks.py]
from random import uniform as ran, choice
L=[(0.25,0.25),(0.75,0.25),(0.25,0.75),(0.75,0.75)]
sigma,delta=0.20,0.15
for iter in range(1000000):
  a = choice(L)
  L.remove(a)
  b = (a[0] + ran(-delta,delta),a[1] + ran(-delta,delta))
  min_dist = min((b[0]-x[0])**2 + (b[1]-x[1])**2 for x in L)
  box_cond = min(b[0],b[1]) < sigma or max(b[0],b[1]) >1-sigma
  if box_cond or min_dist < 4*sigma**2:
   L.append(a)
  else:
   L.append(b)
print L
\end{lstlisting}

\smacfigure{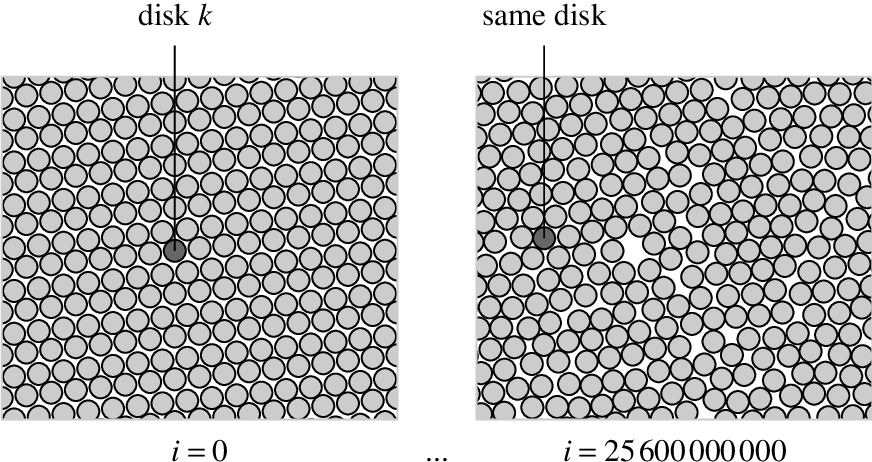} {Initial and final configurations
for $256$ disks at density $\eta = 0.72$ (much smaller than the
close-packing density) after $25.6$ billion iterations. The sample at
iteration $i=25.6\times 10^9$ remembers the initial configuration.}

The Monte-Carlo dynamics of the Markov-chain algorithm,
in \fig{markov_disks_movie},  superficially resembles the
molecular-dynamics in \fig{event_movie_double}. Let us work out the
essential differences between the two: In fact, the Monte Carlo dynamics
is diffusive: It can be described in terms of diffusion constants and
transfer matrices, and convergence towards the equilibrium distribution of
\eq{direct-diskdens} is exponential. In contrast, the dynamics of the
event-chain algorithm is hydrodynamic (it contains eddies, turbulence,
etc, and their characteristic timescales). Although it converges to
the same equilibrium state, as discussed before, this convergence is
algebraic. This has dramatic consequences, especially in two dimensions
(see \SMAC\ Sect. 2.2.5), even though the long-time dynamics in a finite
box is more or less equivalent in the two cases. This is the fascinating
subject of long-time tails, discovered by Alder and Wainwright (1970),
which for lack of time could not be covered in my lectures (see \SMAC\
Sect. 2.2.5).

The local Markov-chain Monte Carlo algorithm runs into serious trouble at
high density, where the Markov chain of configurations effectively gets
stuck during long times (although it remains ergodic). The cleanest
illustration of this fact is obtained by starting the run with
an easily recognizable initial configuration, as the one shown in
\fig{tilted_conf_two}, which is slightly tilted with respect to the
$x$-axis. We see in this example that  $\taucorr \gg 2.56 \times 10^9$
iterations, even though it remains finite and can be measured in a Monte
Carlo calculation.

Presently, no essentially faster algorithm than the local algorithm
is known for uniform hard spheres, and many practical calculations
(done with very large numbers of particles at high density) are
clearly un-converged. Let us notice that the slowdown of the Monte
Carlo calculation at high density has a well-defined physical origin:
the slowdown of the single-particle diffusion at high density. However,
this does not exclude the possibility of much faster algorithms, as we
will discuss in the fourth lecture.

\section{Cluster algorithms, birth-and-death processes}
\slabel{cluster_birth}

In the present section, we explore Monte Carlo algorithms that are not
inspired by the physical process of single-particle diffusion underlying
the local Markov-chain Monte Carlo algorithm.  Instead of moving one
particle after the other, cluster algorithms construct coordinated moves
of several particles at a time, from  one configuration, $a$, to a very
different configuration, $b$ in one deterministic step.  The pivot
cluster algorithm (Dress and Krauth 1995, Krauth and Moessner 2003)
is the simplest representative of a whole class of algorithms.

\smacfigure{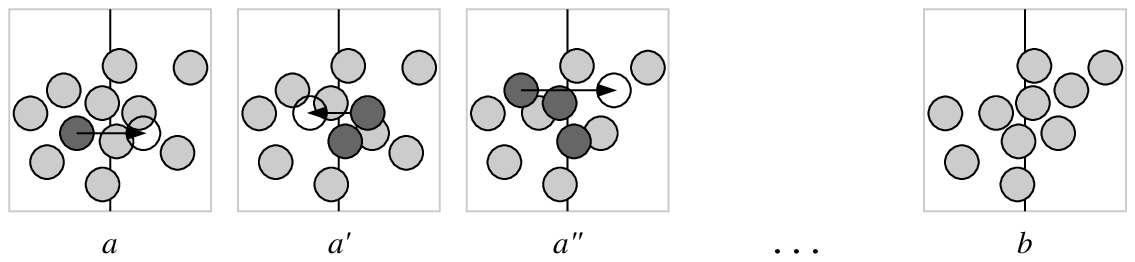}{A cluster move about a pivot axis involving
five disks. The pivot-axis position and its orientation (horizontal,
vertical, diagonal) are chosen randomly. Periodic boundary conditions
allow us to scroll the pivot into the center of the box.}

In this algorithm, one uses a \quot{pocket}, a stack of disks that
eventually have to move.  Initially, the pocket contains a random disk.
As long as there are disks in the pocket, one takes one of them out of it
and moves it. It gets permanently placed, and all particles it overlaps
with are added to the pocket (see \fig{cluster_pocket}, the \quot{pocket
particles} are colored in dark). In order to satisfy detailed balance,
the move $a \to b = T(a)$ must have a $Z_2$ symmetry $a=T^2(a)$ (as a
reflection around a point, an axis, or a hyperplane),  such that moving
it twice brings each particle back to its original position (see \SMAC\
Sect. 2.5.2). In addition, the transformation $T$ must map the simulation
box onto itself. For example, we can use a reflection around a diagonal
in a square or cube box, but not in a rectangle or cuboid. Periodic
boundary conditions are an essential ingredient in this algorithm. In
Python, the pocket algorithm can be implemented in a dozen lines of code
(see \prog{pocket-disks.py}).

\begin{lstlisting}[float,caption=pocket-disks.py in a periodic box of size $1 \times 1$]
from random import uniform as ran, choice
import box_it, dist # see SMAC  web site for periodic distance 
Others=[(0.25,0.25),(0.25,0.75),(0.75,0.25),(0.75,0.75)]
sigma_sq=0.15**2
for iter in range(10000):
   a = choice(Others)
   Others.remove(a)
   Pocket = [a]
   Pivot=(ran(0,1),ran(0,1))
   while Pocket != []:
      a = Pocket.pop()
      a = T(a,Pivot)
      for b in Others[:]: # "Others[:]" is a copy of "Others"
      if dist(a,b) < 4*sigma_sq:
         Others.remove(b)
         Pocket.append(b)
      Others.append(a)
print Others
\end{lstlisting}

\smacfigure{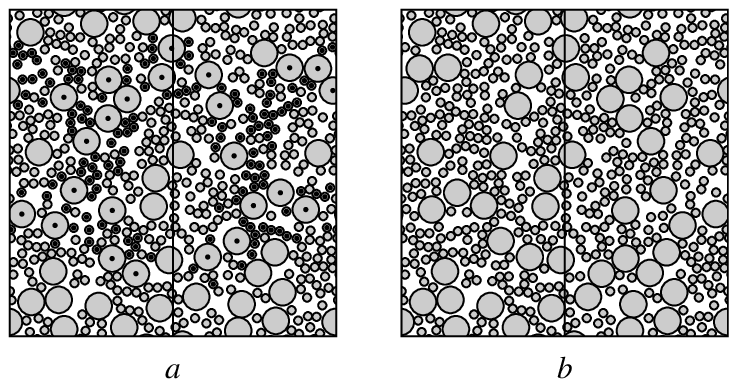} {Single iteration (from \emph{a} to \emph{b})
of the cluster algorithm for a binary mixture of hard spheres. The
symmetry operation is a flip around the $y$ axis shown. Transformed
disks are marked with a dot (see Buhot and Krauth, 1999).  }
The pivot-cluster algorithm fails at high density, say, at the
condition of \fig{tilted_conf_two}, where the transformation $T$
simply transforms all particles in the system. In that case entire
system without changing the relative particle positions. However,
the algorithm is extremely efficient for simulations of monomer--dimer
models or in binary mixtures, among others.  An example of this is given
in \fig{binary_disks}.

At the end of this lecture, let us formulate  hard-sphere systems with
a grand-canonical partition function and fugacity  $\lambda$:
\begin{equation*}
Z = \sum_{N=0}^{\infty} \lambda^N \pi(x_1 \TO x_N).
\elabel{Z_grand_canonical}
\end{equation*}
and discuss the related Markov-chain Monte Carlo algorithms in terms of
birth-and-death processes (no existential connotation intended): Between
two connected configurations, the configuration can only change through
the appearance (\quot{birth}) or disappearance (\quot{death}) of a disk
(see \fig{birth_death}) \smacfigure{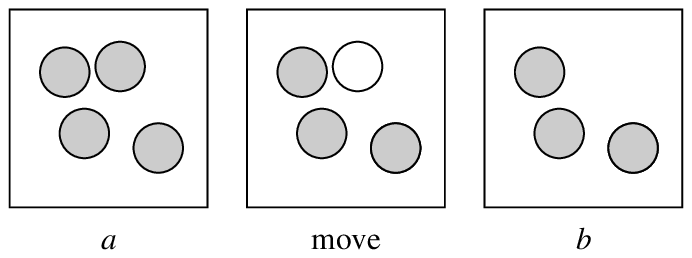} {Birth-and-death process
for grand-canonical hard disks.} It follows from the detailed balance
condition \eq{detailed_balance} that the probability to $\pi(a) = \lambda
\pi(b)$, so that $p(b \to a) = \lambda p(a \to b)$. This means that to
sample \eq{Z_grand_canonical}, we simply have to install one \quot{death}
probability (per time interval $\ddd{t}$) for any particle, as well as
a birth probability for creating a new particle anywhere in the system
(this particle is rejected if it generates an overlap).

As in the so-called \quot{faster-than-the-clock} algorithms (see \SMAC\
Sect. 7.1) one does not discretize time $\tau \to \Deltau$ but rather
samples lifetimes and birth times from their exponential distributions. In
\fig{time_space}, we show a time-space diagram of all the events that
can happen in one extended simulation (as in the first lecture), in
the coupling-from-the-past framework, as used by Kendall and Moller
(2000). We do not know the configuration of disks but, on a closer look,
we can deduce it, starting from the time $t_0$ indicated.

\smacfigure{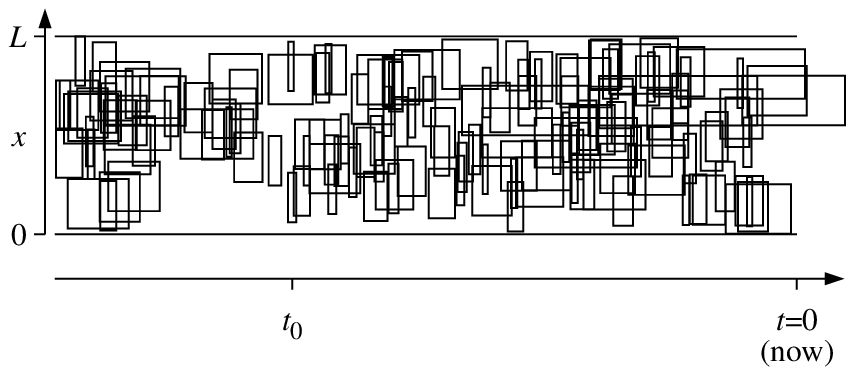} {One-dimensional time-space diagram of
birth-and-death processes for hard disks in one dimension. }

\chapter{Quantum Monte Carlo simulations}

\section{Density matrices, naive quantum simulations}
\slabel{density_matrices}

After the connection between classical mechanics and (classical)
statistical mechanics, we now investigate the junction between statistical
mechanics and quantum physics. We first consider a single particle of
mass $m$ in a harmonic potential $V(x) = \half \omega x^2$, for which
wave functions and energy eigenvalues are all known (see \fig{hermite},
we use units $\hbar = m = \omega=1$).

\smacfigure{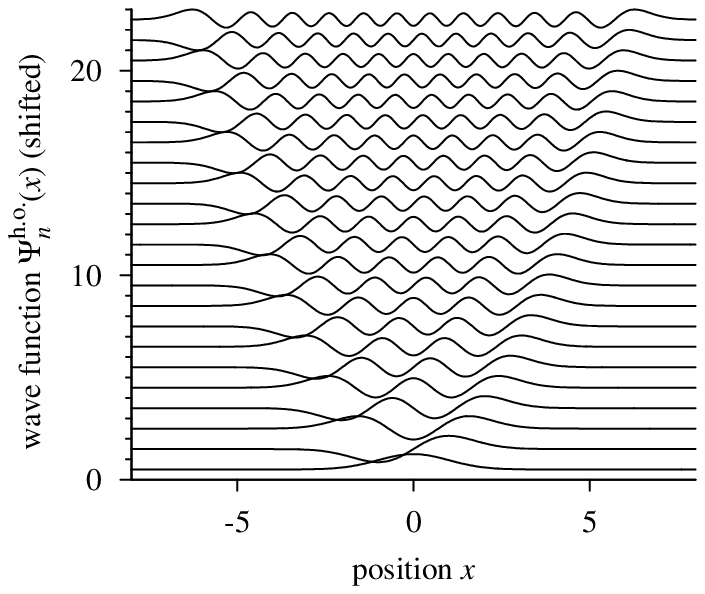}{Harmonic-oscillator wave functions $\psi_n$,
shifted by the energy eigenvalue $E_n$.}
As before, we are interested in the probability $\pi(x)$ for a particle
to be at position $x$ (compare with \eq{direct-diskdens}).  This
probability can be assembled from the statistical weight
of level $k$, $\pi(E_k) \propto \expb{-\beta E_k}$
and from the quantum mechanical probability 
$ \psi_k(x) \psi_k^*(x)$ to be at position $x$ while in level $k$, 
\begin{equation}
\pi(x) = \frac{1}{Z(\beta)}\underbrace{\sum_k \expb{-\beta E_k} 
\Psi_k(x) \Psi^*_k(x)}_{\rhomat{x}{x}{\beta}}.
\elabel{density_matrix_def}
\end{equation}
This probability involves a diagonal element of the 
density matrix given by
$\rhomat{x}{x'}{\beta} =\sum_k \expb{-\beta E_k} \Psi_k(x) \Psi^*_k(x)$,
whose trace is the 
partition function:
\begin{equation*}
Z(\beta)= \int\dd{x} \rhomat{x}{x}{\beta} = \sum _k \expb{-\beta E_k}
\elabel{}
\end{equation*}
(see \SMAC\ Sect 3.1.1). For the free particle and the harmonic
oscillator, we can indeed compute the density matrix exactly (see the
later \eq{density_matrix_free_harm}) but in general, eigenstates or
energies are out of reach for more complicated Hamiltonians.  The path
integral approach obtains the density matrix without knowing the spectrum
of the Hamiltonian, by using the convolution property
\begin{equation}
\rhomat[any]{x}{x'}{\beta} = \int \dd{x''} \rhomat[any]{x}{x''}{\beta'}
\rhomat[any]{x''}{x'}{\beta - \beta'} \quad \text{(convolution)}, 
\elabel{convolution_density_matrix}
\end{equation}
which yields the density matrix at a given temperature through a product
of density matrices at higher temperatures. In the high-temperature limit,
the density matrix for the Hamiltonian $H= H^\free + V$ is given by the
Trotter formula
\begin{equation}
\rhomat[any]{x}{x'}{\beta} \xrightarrow[\beta \rightarrow 0]{} \expa{-\half \beta
V(x)}\rhomat[free]{x}{x'}{\beta}\ \expa{-\half \beta V(x')}\quad \text{(Trotter formula)}.
\elabel{hightemperature}
\end{equation}
(compare with \SMAC\ Sect. 3.1.2).
For concreteness, we continue our discussion with the harmonic potential
$V = \half x^2$, although the discussed methods are completely general.
Using \eq{convolution_density_matrix} repeatedly, one arrives at the path-integral 
representation of the partition function  in terms of high-temperature density matrices:
\begin{multline}
Z(\beta) = \int \dd{x} \rho(x,x,\beta) = \int \dd{x_0} \dots \int \dd{x_{N-1}} \\
\underbrace{\rhomat{x_0}{x_1}{\Deltau} \dots \rhomat{x_{N-1}}{x_0}{\Deltau} }_{\pi(x_0 \TO x_{N-1})}
\elabel{path_integral_Z}
\end{multline}
where $\Deltau=\beta/N$. In the remainder of this lecture we will focus on this  multiple integral
but we are again more interested in sampling (that is, in generating \quot{paths}
$\SET{x_0 \TO x_{N-1}}$ with probability $\pi(x_0 \TO x_{N-1})$) than
in actually computing it.
A naive Monte Carlo sampling algorithm is set up in a few lines of
code (see the \prog{naive-harmonic-path.py}, on the \SMAC\ web site).
In the integrand of \eq{path_integral_Z}, one  chooses a random point $k
\in [0,N-1]$ and a uniform random displacement $\delta_x \in [-\delta,
\delta]$ (compare with \fig{naive_path_moves}). The acceptance
probability of the move depends on the weights $\pi(x_0 \TO x_{N-1})$,
thus both on the free density matrix part, and on the interaction
potential.

\smacfigure{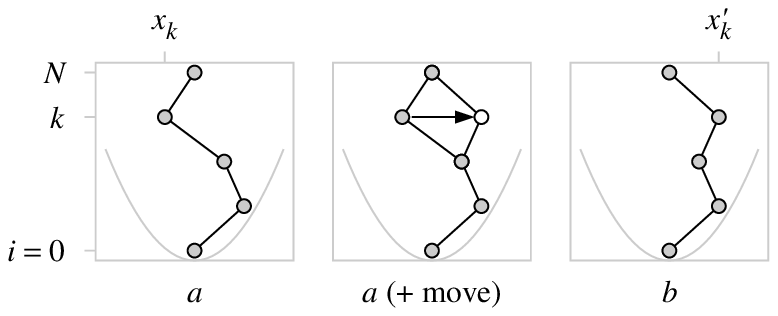}{Move of the path $\SET{x_0 \TO x_{N-1}}$
\emph{via} element $k \in [0,N-1]$, sampling the partition function in
\eq{path_integral_Z}. For $k=0$, both $0$ and $N$ are displaced in the
same direction. The external potential is indicated.}

%\begin{lstlisting}[float,caption=naive-harmonic-path.py]
%from math import pi, exp, sqrt
%from random import uniform as ran, randint as nran
%N,beta,delta=4,2.,0.5
%del_tau=beta/N
%def rho_free(x,xp,beta):
%   return exp(-(x-xp)**2/(2.*beta))
%#------------------------------------------------------------------
%x =[0. for k in range(N)]
%for iter in range(1000000):
%   k = nran(0,N-1)
%   x_old=x[k]
%   x_new=x_old + ran(-delta,delta)
%   xp=x[(k+1)%N]
%   xm=x[k-1]
%   pi_old = rho_free(xm,x_old,del_tau)*\
%      rho_free(x_old,xp,del_tau)*exp(-del_tau*x_old**2/2.)
%   pi_new = rho_free(xm,x_new,del_tau)*\
%      rho_free(x_new,xp,del_tau)*exp(-del_tau*x_new**2/2.)
%   Upsilon=pi_new/pi_old
%   if (ran(0,1) < Upsilon): x[k]=x_new
%   if iter%10 ==0: print x
%\end{lstlisting}

%\smacfigure{naive_harmonic_movie}{qqqqqqq}

The naive algorithm is extremely slow because it moves only a single
\quot{bead} $k$ out of $N$, and not very far from its neighbors $k-1$
and $k+1$.  At the same time, displacements of  several beads at the
same time would rarely be accepted.  In order to go faster through
configuration space, our proposed moves  must learn about quantum
mechanics.  This is what we will teach them in the following section.

\section{Direct sampling of a quantum particle: L\'{e}vy construction}
\slabel{levy_construction}
The problem with naive path-sampling is that the algorithm lacks
insight: the probability distribution of the proposed moves contains
no information about quantum mechanics.  However, this  problem can be
solved completely for a free quantum particle and also for a particle
in a harmonic potential $V(x)= \half x^2$, because in both cases the
exact density matrix for a particle of mass $m=1$, with $\hbar=1$, is a Gaussian:
\begin{equation}
\rhomat{x}{x'}{\beta} = 
\begin{cases}
\frac{1}{\sqrt{2 \pi\beta}} \expc{-\frac{(x-x')^2}{2 \beta}}\quad \text{(free particle)} \\
\frac{1}{\sqrt{2 \pi\sinh \beta}} \expc{
   -\frac{(x+x')^2}{4}\tanh\frac{\beta}{2} 
   -\frac{(x-x')^2}{4}\coth\frac{\beta}{2} 
}
\quad \text{(osc.)}
\end{cases}
\elabel{density_matrix_free_harm}
\end{equation}  

The distribution of an intermediate point, for example in the left panel
of \fig{levy_schema}, is given by
\begin{equation*}
\pi(x_1|x_0,x_N)=\underbrace{\rhomat{x_0}{x_1}{\frac{\beta}{N}}}_{\text{Gaussian, see \eq{density_matrix_free_harm}}}
\underbrace{\rhomat{x_1}{x_N}{\frac{(N-1)\beta}{N}}}_{\text{Gaussian, see \eq{density_matrix_free_harm}}}.
\end{equation*}
As a product of two Gaussians, this is again a Gaussian, and it can be
sampled directly. After sampling $x_1$, one can go on to $x_2$, etc.,
until the the whole path is constructed, and without rejections
(L\'{e}vy 1940).

\smacfigure{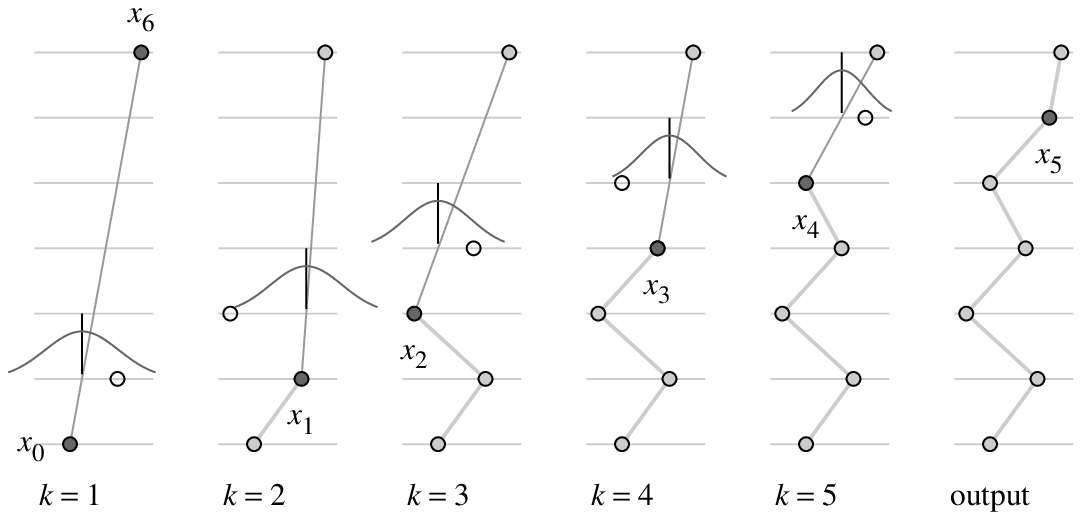}{L\'{e}vy construction of a path contributing
to the free-particle density matrix $\rhomat{x_0}{x_6}{\beta}$. The
intermediate distributions, all Gaussians, can be sampled directly.}

\begin{lstlisting}[float,caption=levy-free-path.py]
from math import pi, exp, sqrt
from random import gauss
N = 10000
beta = 1.
Del_tau = beta/N
x = [0 for k in range(N+1)]
y = range(N+1)
for k in range(1,N):
  Del_p = (N-k)*Del_tau
  x_mean = (Del_p*x[k-1] + Del_tau*x[N])/(Del_tau + Del_p)
  sigma = sqrt(1./(1./Del_tau + 1./Del_p))
  x[k] = gauss(x_mean,sigma)
print x
\end{lstlisting}
The L\'{e}vy construction is exact for a free particle in a  harmonic
potential and of course also for free particles (as shown in the Python code). 
Direct-sampling algorithms can be tailored to specific
situations, such as periodic boundary conditions or hard walls (see \SMAC\
Sections  3.3.2 and 3.3.3).  Moreover, even for generic interactions,
the L\'{e}vy construction allows to pull out, treat exactly, and
sample without rejections the free-particle Hamiltonian. In the generic
Hamiltonian $H= H_0 + V$, the Metropolis rejection then only takes care of
the interaction term $V$.  Much larger moves than before become possible
and the phase space is run through more efficiently than in the naive
algorithm, where the Metropolis rejection concerned the entire $H$.
This makes possible nontrivial simulations with a very large number of
interacting particles (see Krauth 1996, Holzmann and Krauth 2008).  We will illustrate
this point in the following section, showing that $N$-body simulations
of ideal (non-interacting) bosons (of arbitrary size) can be done by
direct sampling without any rejection.

We note that the density matrices in \eq{density_matrix_free_harm},
for a single particle in a $d$-dimensional harmonic oscillator yield
the partition functions
\begin{equation}
z^\harm(\beta) =  \int \ddd{x} \rhomat{x}{x}{\beta} = 
\glc 1 - \expb{-\beta}\grc^{-d} \quad \text{(with $E_0=0$)}
\elabel{single_particle_z}
\end{equation}
where we use the lowercase symbol $z(\beta)$ in order to differentiate the
one-particle partition function from its $N$-body counterpart $Z(\beta)$
which we will need in the next section.

\section{Ideal bosons: Landsberg recursion and direct sampling}
\slabel{landsberg_sampling}
The density matrix for $N$ distinguishable particles $\rhomat[dist]{\SET{x_1
\TO x_N}}{\SET{x_1' \TO x_N'}}{\beta}$ is assembled from normalized $N$-particle
wavefunctions and their associated energies, as in \eq{density_matrix_def}.  The density
matrix for indistinguishable particles is then obtained by symmetrizing the
density matrix for $N$ distinguishable particles. This can be done either 
by using symmetrized (and normalized) wavefunctions to construct the density matrix or,
equivalently, by averaging the distinguishable-particle density matrix over
all  $N!$ permutations (Feynman 1972):
\begin{equation}
 Z = \frac{1}{N!}\sum_{P} \int \ddd{^N x}
 \rhomat[dist]{\SET{x_1 \TO x_N}}{\SET{x_{P(1)} \TO x_{P(N)}}}{\beta}.
 \elabel{z_permutation}
\end{equation}
This equation is illustrated in \fig{all_permutations} for four
particles. In each diagram, we arrange the points $x_1,x_2,x_3,x_4$
from left to right and indicate the permutation by lines.  The final
permutation (to the lower right of \fig{all_permutations}), corresponds
for example to the permutation $P(1,2,3,4)=(4,3,2,1)$.  It consists of
two cycles of length $2$, because $P^2(1)=P(4)=1$ and $P^2(2)=P(3)=2$.

\smacfigure{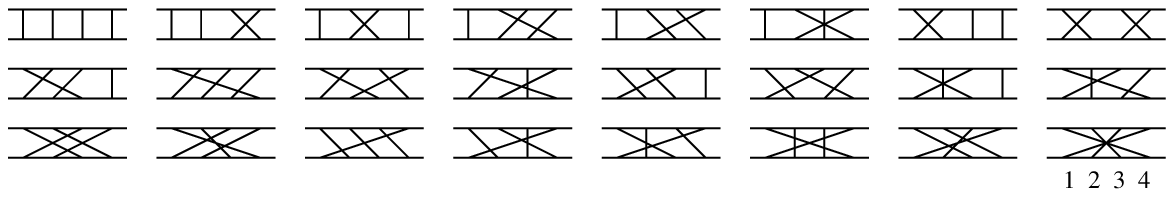}{The $24$
permutations contributing to the partition function for 4 ideal bosons.}

To illustrate \eq{z_permutation}, let us compute the contribution to
the partition function stemming from this permutation for free particles:
\begin{multline*}
Z_{4321}= \int\ddd{x_1} \ddd{x_2} \ddd{x_3} \ddd{x_4}
 \rhomat[free]{\SET{x_1,x_2,x_3,x_4}}{\SET{x_4, x_3, x_2, x_1}}{\beta}=\\
\underbrace{\glc \int \ddd{x_1} \ddd{x_4} \rhomat[free]{x_1}{x_{4}}{\beta} 
 \rhomat[free]{x_4}{x_1}{\beta}\grc}_{\int\ddd{x_1} \rhomat[free]{x_1}{x_1}{2\beta}= 
 z(2 \beta)\quad \text{see \eq{convolution_density_matrix}}}
\glc \int \ddd{x_2}\ddd{x_3} \rhomat[free]{x_2}{x_3}{\beta}
\rhomat[free]{x_3}{x_2}{\beta}
\grc.
\end{multline*}
This partial partition function of a four-particle system
writes as the product of one-particle partition functions. The number
of terms corresponds to the number of cycles and the length of cycles
determines the effective inverse temperatures in the systems (Feynman
1972).

The partition function for $4$ bosons is given by a sum over $24$
terms, $5$ bosons involve $120$ terms, and the number of terms in
the partition function for a million bosons has more than 5 million
digits. Nevertheless, this sum over permutations can be computed exactly 
through an ingenious recursive procedure of $N$ steps due to Landsberg (1961)
(see \fig{all_permutations_landsberg}).

\smacfigure{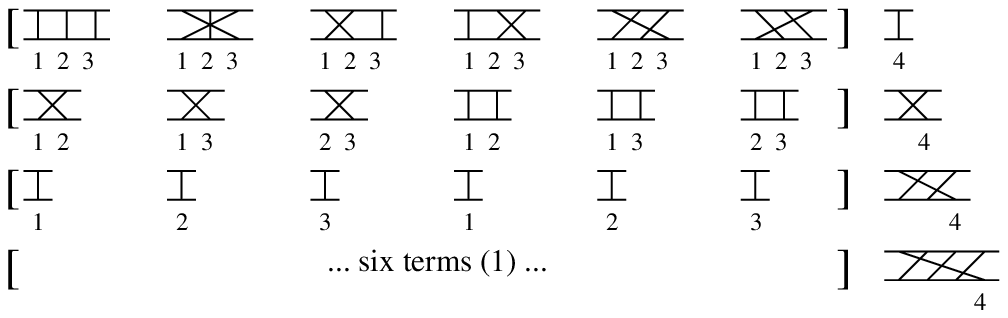}{The $24$ permutations of \fig{all_permutations},
with the last-element cycle pulled out of the diagram.}

\figg{all_permutations_landsberg} contains the same permutations
as \fig{all_permutations}, but they have been rearranged and the
last-element cycle (the one containing the particle $4$) has been pulled
out. 
These pulled-out terms outside braces make up
the partition function of a single particle at temperature $\beta$
(on the first row), at inverse temperature $2 \beta$ (on the second row),
etc., as already computed in \eq{single_particle_z}. 
The diagrams within braces in \fig{all_permutations_landsberg}  contain the $3!=6$
three-boson partition functions making up $Z_3$ (on the first line of
the figure). The second row contains three times the diagrams making
up $Z_2$, etc.  All these terms yield together the terms in \eq{landsberg_recursion}:
\begin{equation}
Z_N = \frac{1}{N} \glb z_1 Z_{N-1} \PLUSPLUS z_k Z_{N-k} 
      \PLUSPLUS z_N Z_0 \grb \quad (\text{ideal Bose gas})
\elabel{landsberg_recursion}
\end{equation}
(with $Z_0=1$, see \SMAC\ Sect. 4.2.3 for a proper derivation). $Z_0$,
as well as the single-particle partition functions are known from
\eq{single_particle_z}, so that we can first determine $Z_1$, then $Z_2$,
and so on (see the \prog{harmonic-recursion.py}, the \SMAC\ web site contains a version with
graphics output).
\begin{lstlisting}[float,caption=canonic-recursion.py] 
import math
def z(beta,k):
   sum = 1/(1- math.exp(-k*beta))**3
return sum
N=1000
beta=1./5
Z=[1.]
for M in range(1,N+1):
    Z.append(sum(Z[k] * z(beta,M-k) for k in range(M))/M)
pi_list=[z(beta,k)*Z[N-k]/Z[N]/N for k  in range(1,N+1)]
\end{lstlisting}

The term in the Landsberg relation of \eq{landsberg_recursion} $\propto
z_k Z_{N-k}$ can be interpreted as a cycle weight, the statistical weight
of all permutations in which the particle $N$ is in a cycle of length $k$
\begin{equation*}
\pi_k = \frac{1}{N Z_N} z_k Z_{N-k} \quad \text{(cycle weight)}.
\end{equation*}
From the Landsberg recursion, we can explicitly compute cycle
weights for arbitrary large ideal-boson systems at any temperature
(see \fig{cycle_distribution}).

\smacfigure{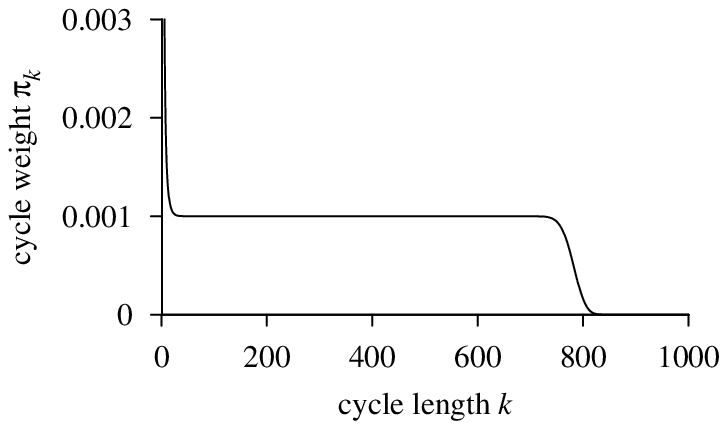}{Cycle-weight distribution for $1000$
ideal bosons in a three-dimensional harmonic trap, obtained from
\prog{canonic-recursion.py}, exactly as written}

In \fig{cycle_distribution}, we notice that the cycle-weight distribution
$\pi_k$ is flat for most $k$ before it drops to zero around $k \sim
780$. Curiously, the derivative of this function yields the distribution
of the condensate fraction (Holzmann and Krauth 1999, Chevallier and
Krauth 2007), so that we see that at the temperature chosen, there are
about $780$ particles in the groundstate ($N_0/N \simeq 0.78$). At
higher temperatures, the distribution of cycle weights is narrower.
This means that there are no long cycles.

We now turn the \eq{landsberg_recursion} around, as we first did for
the Gaussian integral: rather than computing $Z(\beta)$
from the cycle weights, we sample the cycle distribution from its weights that is, 
pick one of the $\pi_k$ with 
\begin{equation*}
\pi_1 \TO \pi_k \TO \pi_N.
\elabel{}
\end{equation*}
(we pick a cycle length $1$ with probability $\pi_1$, $2$ with
probability $\pi_2$, and so on (it is best to use the tower-sampling
algorithm we mentioned in the first lecture). Suppose we sampled a
cycle length $k$. We then know that our Bose gas contains a cycle
of length $k$, and we can sample the three-dimensional positions of
the $k$ particles on this cycle from the three-dimensional version of
\prog{harmonic-levy.py} for $k$ particles instead of $N$, and at inverse
temperature $k \beta$. Thereafter, we sample the next cycle length
from the Landsberg recursion relation with $N-k$ particles instead of
$N$, and so on, until all particles are used up (see the Appendix for a 
complete program in $44$ lines). Output of this
program is shown in \fig{direct_harmonic_sample}, projected onto two
dimensions. As in a real experiment, the three-dimensional harmonic
potential confines the particles but, as we pass the Bose--Einstein
transition temperature (roughly at the temperature of the left panel of
\fig{direct_harmonic_sample}), they start to move to the center and
to produce the landmark peak in the spatial density.  At the temperature
of the right-side panel, roughly $80 \%$ of particles are condensed into
the ground state.

\smacfigure{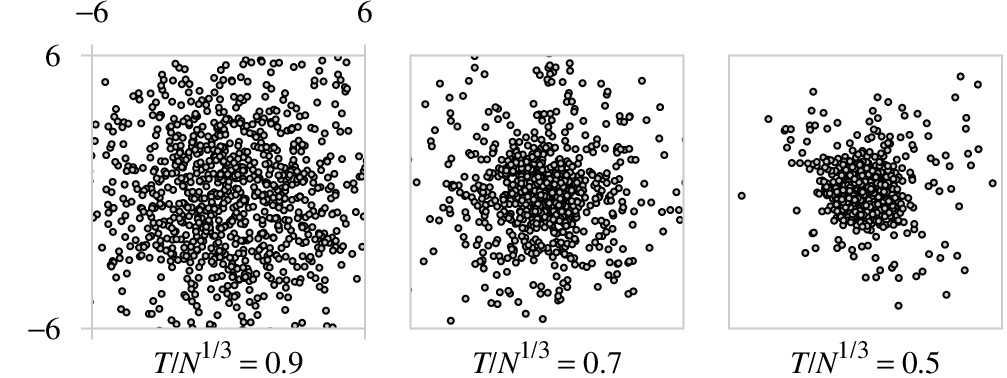}{Two-dimensional snapshots of
$1000$ ideal bosons in a three-dimensional harmonic trap (obtained with
the direct-sampling algorithm).}

The power of the path-integral approach resides in the facility with which
interactions can be included. This goes beyond what can be treated in an
introductory lecture (see \SMAC\ Sect. 3.4.2 for an in-depth discussion).
For the time being we should take pride in our rudimentary sampling
algorithm for ideal bosons, a true quantum Monte Carlo program in
a nutshell. 

\chapter{Spin systems: samples and exact solutions}
\section{Ising Markov-chains: local moves, cluster moves}
\slabel{ising_markov}
In this final lecture, we study models of discrete spins $\SET{\sigma_1
\TO \sigma_N}$  with $\sigma_k = \pm 1$ on a lattice with $N$ sites, with energy
\begin{equation}
E    = -  \sum_{\neigh{k}{l}}J_{kl}
       \s_k \s_l.
\elabel{ising_pair_energy}
\end{equation}
Each pair of neighboring sites $k$ and $l$ is counted only once. In
\eq{ising_pair_energy}, we may choose all the $J_{kl}$ equal to
$+1$. We then have the ferromagnetic Ising model. If we choose random
values $J_{kl} = J_{lk}= \pm1$, one speaks of the Edwards--Anderson
spin glass model. Together with the hard-sphere systems, these models
belong to the hall of fame of statistical mechanics, and have been
the crystallization points for many developments in computational
physics. Our goal in this lecture will be two-fold. We shall illustrate
several algorithms for simulating Ising models and Ising spin glasses
in a concrete setting. We shall also explore the relationship between
the Monte Carlo sampling approach and analytic solutions, in this case
the analytic solution for the two-dimensional Ising model initiated by
Onsager (1942), Kac and Ward (1952), and Kaufmann (1949).

\smacfigure{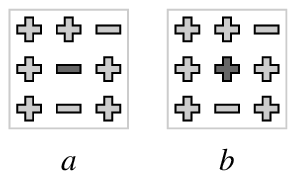} {A spin flip in the Ising
model. Configuration $a$ has central field $h=2$ (three \quot{$+$}
neighbors and one \quot{$-$} neighbor), and configuration $b$ has $h=-2$. The
statistical weights satisfy $\pi_b/\pi_a = \expb{-4\beta}$.}

\begin{lstlisting}[float,caption=markov-ising.py] 
from random import uniform as ran, randint,choice 
from math import exp 
import square_neighbors # defines the neighbors of a site 
L = 32 
N = L*L 
S = [choice([-1,1]) for k in range(N)] 
beta = 0.42 
nbr = square_neighbors(L) 
# nbr[k] = (right,up,left,down) 
for i_sweep in range(100):
   for iter in range(N):
      k = randint(0,N-1) 
      h = sum(S[nbr[k][j]] for j in range(4))
      Delta_E = 2.*h*S[k] 
      Upsilon = exp(-beta*Delta_E) 
      if ran(0.,1.) < Upsilon: S[k] = -S[k]
print S
\end{lstlisting}

The simplest Monte Carlo algorithm for sampling the partition function
\begin{equation*}
Z = \sum_{\text{confs $\sigmavec$}}\expc{-\beta E(\sigmavec)}
\elabel{}
\end{equation*}
picks a site at random, for example the central site on the square
lattice of configuration $a$ in \fig{ising_3x3_spin_flip}. Flipping
this spin would produce the configuration $b$. To satisfy detailed
balance, \eq{detailed_balance}, we must accept the move $a \to b$
with probability $\min(1, \pi_b/\pi_a)$, as implemented in the program
\prog{markov-ising.py} (see \SMAC\ Sect. 5.2.1). This algorithm is very
slow, especially near the phase transition between the low-temperature
ferromagnetic phase and the high-temperature paramagnet.  This slowdown
is due to the fact that, close to the transitions, configurations with
very different values of the total magnetization $M = \sum_k \s_k$
contribute with considerable weight to the partition function (in two
dimension, the distribution ranges practically from $-N$ to $N$). One
step of the local algorithm changes the total magnetization at most by
a tiny amount, $\pm 2$, and it thus takes approximately $N^2$ steps (as
in a random walk) to go from one configuration to an independent one.
This, in a nutshell,  is the phenomenon of critical slowing down.

One can overcome critical slowing down by flipping a whole cluster of
spins simultaneously, using moves that know about statistical mechanics
(Wolff 1989).  It is best to start from a random site, and then to
repeatedly add to the cluster, with probability $p$, the neighbors of
sites already present if the spins all have the same sign. At the end
of the construction, the whole cluster is flipped. We now compute the
value of $p$ for which this algorithm has no rejections (see \SMAC\
Sect. 5.2.3).

\smacfigure{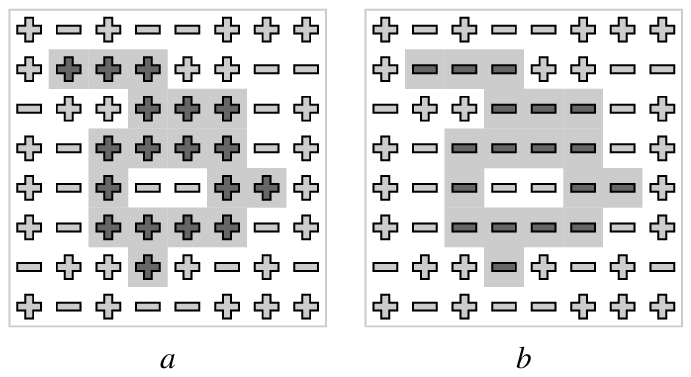}
{A large cluster of like spins in the Ising model. The construction stops
with the gray cluster in $a$ with probability $(1-p)^{14}$, corresponding
to the $14$ \quot{$++$} links across the boundary.  The corresponding
probability for  $b$ is $(1-p)^{18}$.}

The cluster construction stops with the configuration $a$
of \fig{spins_wolff} with probability $p(a \to b) = \const
(1-p)^{14}$, one factor of $(1-p)$ for every link \quot{$++$} across
the cluster boundary.  Likewise, the cluster construction in $b$ stops
as shown in  $b$ if the $18$
neighbors \quot{$--$} were considered without success (this happens
with probability $p(b \to a) = \const (1-p)^{18}$). The detailed balance
condition relates the construction probabilities to the Boltzmann weights
of the configurations
\begin{align*}
\pi_a &= \const'\expc{-\beta(-14+18)}\\
\pi_b &= \const' \expc{-\beta(-18+14)}, 
\elabel{}
\end{align*}
where the $\const'$ describes all the contributions to the energy not coming from the 
boundary. We may enter stopping probabilities and Boltzmann weights
into the detailed balance condition $\pi_a p(a \to b) = \pi_b \pi(b \to a)$
and find
\begin{equation}
\expb{14 \beta} \expb{- 18 \beta} (1-p)^{14} = \expb{- 14 \beta} \expb{18 \beta} (1-p)^{18}.
\elabel{detailed_balance}
\end{equation}
This is true for $p=1- \expb{-2\beta}$, and is independent
of our example, with its $14$ \quot{$++$} and $18$ neighbors \quot{$--$}
links (see \SMAC\ Sect. 4.2.3).

The cluster algorithm is implemented in a few lines (see
\prog{cluster-ising.py}) using the pocket approach of \sect{cluster_birth}: Let the
\quot{pocket} comprise those sites of the cluster whose neighbors have
not already been scrutinized. The algorithm starts by putting a random site
both into the cluster and the pocket.  One then takes a site out of the
pocket, and adds neighboring sites (to the cluster and to the pocket)
with probability $p$ if their spins are the same, and if they are not
already in the cluster.  The construction ends when the pocket is empty. We
then flip the cluster.

\smacfigure{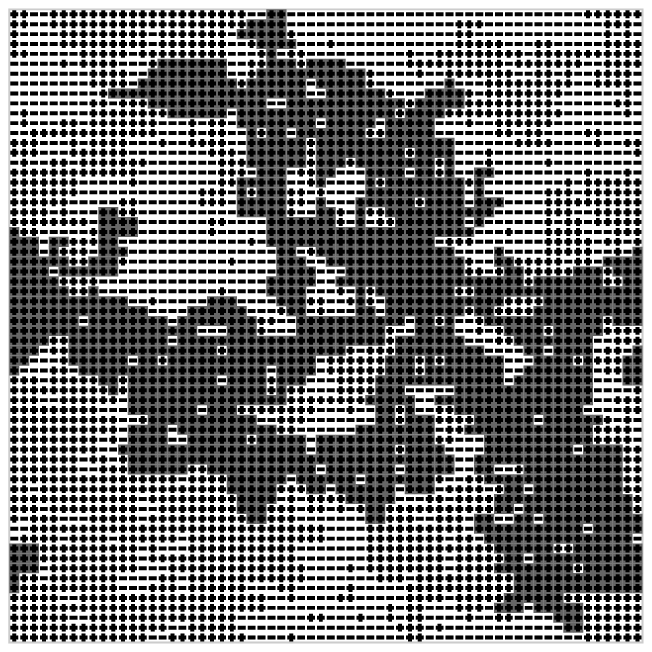}{A large cluster with $1548$ spins in a
$64\times 64$ Ising model with periodic boundary conditions. All the
spins in the cluster flip together from \quot{$+$} to \quot{$-$}.}

\begin{lstlisting}[float,caption=cluster-ising.py]
from random import uniform as ran, randint,choice
from math import exp
import square neighbors # defines the neighbors of a site  
L = 32
N = L*L
S = [choice([-1,1]) for k in range(N)]
beta = 0.4407
p = 1-exp(-2*beta)
nbr = square_neighbors(L) # nbr[k]= (right,up,left,down)
for iter in range(100):
   k = randint(0,N-1)
   Pocket = [k]
   Cluster = [k]
   while Pocket != []:
      k = choice(Pocket)
      for l in nbr[k]:
         if S[l] == S[k] and l not in Cluster and ran(0,1) < p:
            Pocket.append(l)
            Cluster.append(l)
      Pocket.remove(k)
   for k in Cluster: S[k] = -S[k]
print S
\end{lstlisting}
Cluster methods play a crucial role in computational statistical
physics because, unlike local algorithms and unlike experiments, they
do not suffer from  critical slowing down.  These methods have spread
from the Ising model to many other fields of statistical physics.

Today, the non-intuitive rules for the cluster construction are well
understood, and the algorithms are very simple.  In addition, the modern
meta languages are so powerful that a rainy Les Houches afternoon provides
ample time to implement the method, even for a complete non-expert in
the field.

\section{Perfect sampling: semi-order and patches} 
\slabel{perfect_sampling_semi_order}
The local Metropolis algorithm picks a site at random and flips it with
the probability $\min(1,\pi_b/\pi_a)$ (as in \sect{ising_markov}).
An alternative local Monte Carlo scheme is the heat-bath
algorithm, where the spin is equilibrated in its local environment
(see \fig{ising_3x3_heat_bath_explained}).  This means that in the
presence of a molecular field $h$ at site $k$, the spin points up and
down with probabilities $\pi^{+}_h$ and $\pi^{-}_h$, respectively, where
\begin{equation}
\begin{split}
\pi^{+}_{h}  & = \frac{\expa{-\beta E^{+}}}{\expa{-\beta E^{+}} + \expa{ - \beta E^{-}}}
 = \frac{1}{1 + \expa{ -2 \beta h}},  \\
\pi^{-}_{h}& = \frac{\expa{-\beta E^{-}}}{\expa{-\beta E^{+}} + \expa{ - \beta E^{-}}}
 = \frac{ 1}{1 + \expa{+2 \beta h}}.
\end{split}
\elabel{heat_bath_hplus}
\end{equation}
\smacfigure{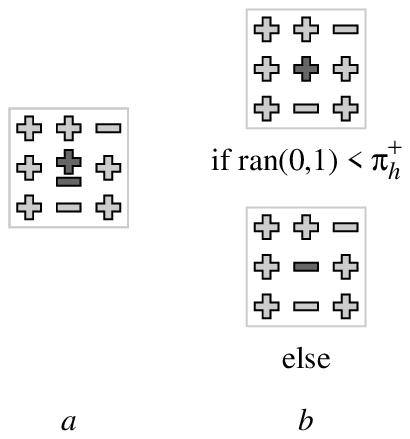}{Heat bath algorithm for the
Ising model. The new position of the spin $k$ (in configuration $b$)
is independent of its original position (in $a$).}

The heatbath algorithm (which is again much slower than
the cluster algorithm, especially near the critical point)
couples just like our trivial simulation in the five-site model of
\sect{Markov_chain_sampling}: At each step, we pick a random site $k$,
and a random number $\Upsilon = \ran{0}{1}$, and apply the Monte Carlo
update of \fig{ising_3x3_heat_bath_explained} with the same $k$ and
the same $\Upsilon$ to all the configurations of the Ising model or
spin glass. After a time $\taucoup$, all input  configurations yield
identical output (Propp and Wilson 1996).

For the Ising model (but not for the spin glass) it is very
easy to compute the coupling time, because the half-order among
spin configurations is preserved by the heat-bath dynamics (see
\fig{ising_3x3_half_order}): we say that a configuration $\sigmavec =
\SET{\sigma_1 \TO \sigma_N}$ is smaller than another configuration
$\sigmavec' = \SET{\sigma_1' \TO \sigma_N'}$ if for all $k$ we have
$\sigma_k \le \sigma_k'$. \footnote{This is a half-order because not
all configurations can be compared to each other.} For the Ising model,
the heat-bath algorithm preserves the half-order between configurations
upon update because a configuration which is smaller than another one
has a smaller field on all sites, thus a smaller value of $\pi_h^+$. We
just have to start the simulation from the all plus-polarized and the
all minus-polarized configurations and wait until these two extremal
configurations couple. This determines the coupling time for all
$2^N$ configurations, exactly as in the earlier trivial example of
\fig{one_d_alternating}.

\smacfigure{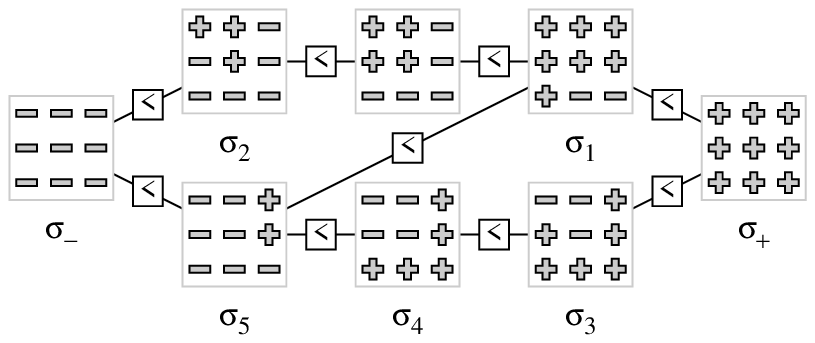}{Half-order in spin models: the
configuration $\sigmavec_-$ is smaller and $\sigmavec_+$ is larger than
all other configurations, which are not all related among each other.  }

The Ising spin glass does not allow such simple tricks to be
played, and we must explicitly check  the coupling for all $2^N$
initial configurations. This enormous surveying task can be simplified
considerably by breaking up the configurations on the entire lattice into
smaller \quot{patches} (see \fig{exact_fig3}). For a given patch size
$M$ ($M=16$ in \fig{exact_fig3}), the total number of configurations
on $N$ patches is bounded by $N 2^M  \ll 2^N$. We can now follow the
heatbath simulation on all individual patches and at the end assemble
configurations on the whole lattice in the  same way as we assemble an
entire puzzle picture from the individual pieces. The procedure can be
made practical, and programmed very easily in modern meta languages such
as Python.

From a more fundamental point of view, the coupling concept in spin
glass models is of interest because the physical understanding of
these systems has seriously suffered from longstanding doubts about
the quality of Monte Carlo calculations thus, at the most basic level,
about the correct calculations of the correlation time.

\smacfigure{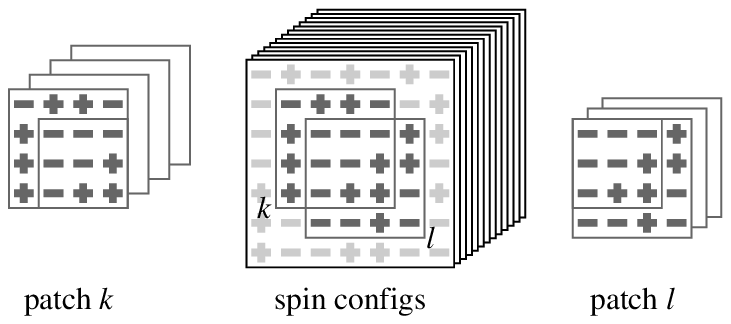}{The very large configuration space of an Ising
spin glass  on $N$ sites viewed from the vantage point of $N$ patches of
smaller size $M$.}

\section{Direct sampling, and the Onsager solution}
\slabel{direct_sampling_onsager}

The two-dimensional Ising model is exactly solvable, as shown by
Onsager (1942): we can compute its critical temperature, and many
critical exponents.  Let us be more precise: The partition function of
the two-dimensional Ising model or the spin glass\footnote{we speak here
of one single \quot{sample} of the spin glass, that is, a given choice of
the couplings $\SET{J_{kl}}$. To compute the average over all couplings is
another  matter.} on a planar lattice with $N$ sites can be expressed as
the square root of the determinant of a $4N\times 4N$ matrix (see \SMAC\
Sect. 5.1.4 for a practical algorithm).  Periodic boundary conditions
can also be handled, it gives four $4N \times 4N$ matrices. This was
used very successfully by Saul and Kardar (1992).  For the Ising model on a finite square
lattice,  these matrices can be diagonalized analytically. This is the
famous analytic expression for $Z(\beta)$  of Kaufman (1949) (see \SMAC\
exerc. 5.9, p. 265).

The solution of the Ising model amounts to summing its high-temperature
series, and this solves an enumeration problem, as evidenced in the
combinatorial solution by Kac and Ward (1949). Indeed, the density
of states of the energy $\NCAL(E)$ can be extracted from the analytic
solution (see Beale (1996)), this means that we can obtain the (integer)
number of states for any energy $E$ for two-dimensional Ising spin glasses
of very large sizes.  The exact solution of the Ising model thus performs
an enumeration, but it counts configurations, it cannot list them.
The final point we elaborate on in these lectures is that, unable to
list configurations, we can still sample them.

The subtle difference between counting and listing of sets implies that
while we access without any trouble the density of states $\NCAL(E)$ we
cannot obtain the distribution (the histogram) of magnetizations and know
very little about the joint distribution of energies and magnetizations,
$\NCAL(M,E)$. Doing so would in fact allow us to solve the Ising model
in a magnetic field, which is not possible.

Let us now expose a sampling algorithm (Chanal and Krauth 2009), which
uses  the analytic solution of the Ising model to compute
$\NCAL(M,E)$ (with only statistical, and no systematic errors) even
though that is impossible to do (exactly). This algorithm constructs the
sample one site after another.  Let us suppose that the gray spins in the
left panel of \fig{onsager_direct} are already fixed, as shown. We can
now set a fictitious coupling $J_{ll}^*$ either to $-\infty$ or two $+
\infty$ and recalculate the partition function $Z_\pm$ with them. The
statistical weight of all configurations in the original partition
function with spin \quot{$+$} is then given by
\begin{equation}
\pi_+ = \frac{Z_+ \expb{\beta J_{kl}}}
{Z_+ \expb{\beta J_{kl}} + Z_- \expb{-\beta J_{kl} }}.
\elabel{direct_sampling}
\end{equation}
and this two-valued distribution can be sampled with one random number.
\eqq{direct_sampling} resembles the heatbath algorithm of \eq{heat_bath_hplus}, but it is not
part of a Markov chain:  After obtaining the value of the spin on site $k$,
we keep the fictitious coupling, and add more sites.  Going over all
sites, we can generate direct samples of the partition function of \eq{},
at any temperature, and with a fixed, temperature-independent effort,
both for the two-dimensional Ising model and the Ising spin glass. This
allows us to obtain arbitrary correlation functions, the generalized
density of states $\NCAL(M, E)$, etc.  We can also use this solution
to determine the behavior of the Ising model in a magnetic field,
which we cannot obtain from the analytic solution, although we use it
in the algorithm.

\smacfigure{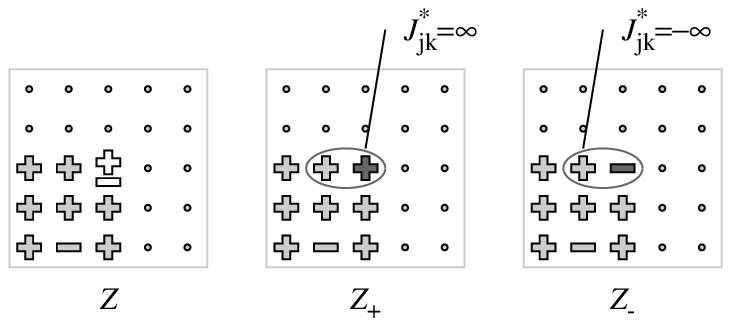}{One iteration in the direct-sampling algorithm
for the two-dimensional Ising model. The probabilities $\pi(\sigma_k=+1)$
and $\pi(\sigma_k=-1)$ (with $k$ the central spin) are obtained from
the exact solution of the Ising model with fictitious couplings
$J_{jk}^*= \pm \infty$.}

The correlation-free direct-sampling algorithm is primarily of theoretical
interest, as it is restricted to two-dimensional Ising model and the Ising
spin glass, which are already well understood. For example, it is known
that the two-dimensional spin glass does not have a finite transition
temperature. Nevertheless, it is intriguing that one can obtain, from
the analytical solution, exactly and  with a performance guarantee,
quantities that the analytical solution cannot give.  The \quot{sampling}
(even the very well controlled, full-performance-guarantee direct sampling
used here) does not meet the limitations of complete enumerations.

\acknowledgements
I would like to thank the organizers of this School for giving me the
opportunity to lecture about some of my favorite subjects.  Thanks to
C. Laumann for introducing me to Python programming and for pointing out
the similarities with \SMAC\ pseudocode and to C.~Chanal and M.~Chevallier
for a careful reading of the manuscript.

\appendix
\chapter{}
Here is an entire Quantum Monte Carlo program (in Python 2.5)) for ideal
bosons in a harmonic trap (see the \SMAC\ web site for a program with
graphics output).
\begin{lstlisting}[caption=direct-harmonic-bosons.py]
from math import sqrt, sinh, tanh, exp
from random import uniform as ran, gauss
def z(beta,k):
   sum = 1/(1-exp(-k*beta))**3
   return sum
def canonic_recursion(beta,N):
   Z=[1.]
   for M in range(1,N+1):
      Z.append(sum(Z[k]*z(beta,M-k) for k in range(M))/M)
   return Z
def pi_list_make(Z,M):
   pi_list=[0]+[z(beta,k)*Z[M-k]/Z[M]/M for k in range(1,M+1)]
   pi_sum=[0]
   for k in range(1,M+1):
      pi_sum.append(pi_sum[k-1]+pi_list[k])
   return pi_sum
def tower_sample(data,Upsilon): #naive, cf. SMAC Sect. 1.2.3
   for k in range(len(data)):
      if Upsilon<data[k]: break
   return k
def levy_harmonic_path(Del_tau,N): #
   beta=N*Del_tau
   xN=gauss(0.,1./sqrt(2*tanh(beta/2.))) 
   x=[xN]
   for k in range(1,N):
      Upsilon_1 = 1./tanh(Del_tau)+1./tanh((N-k)*Del_tau)
      Upsilon_2 = x[k-1]/sinh(Del_tau)+xN/sinh((N-k)*Del_tau)
      x_mean=Upsilon_2/Upsilon_1
      sigma=1./sqrt(Upsilon_1)
      x.append(gauss(x_mean,sigma))
   return x
N=512
beta=1./2.4
Z=canonic_recursion(beta,N)
M=N
x_config=[]
y_config=[]
while M > 0:
   pi_sum=pi_list_make(Z,M)
   Upsilon=ran(0,1)
   k=tower_sample(pi_sum,Upsilon) 
   x_config+=levy_harmonic_path(beta,k)
   y_config+=levy_harmonic_path(beta,k)
   M -= k
\end{lstlisting}

\end{document}